\documentclass[lettersize,journal]{IEEEtran}
\usepackage{amsmath,amsfonts}
\usepackage{algorithmic}
\usepackage{algorithm}
\usepackage{array}
\usepackage[caption=false,font=normalsize,labelfont=sf,textfont=sf]{subfig}
\usepackage{textcomp}
\usepackage{stfloats}
\usepackage{url}
\usepackage{verbatim}
\usepackage{graphicx}
\usepackage{cite}
\usepackage{booktabs}    % professional-quality tables
\usepackage{hyperref}
\DeclareMathOperator*{\argmin}{arg\,min}
\newcommand{\orcid}[1]{\href{https://orcid.org/#1}{\includegraphics[width=8pt]{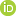}}}
\begin{document}

\title{DiffProsody: Diffusion-based Latent Prosody Generation for Expressive Speech Synthesis with Prosody Conditional Adversarial Training}

\author{Hyung-Seok Oh\orcid{0000-0001-7229-8123}, 
    Sang-Hoon Lee\orcid{0000-0002-8925-4474}, 
    and Seong-Whan Lee\orcid{0000-0002-6249-4996}, ~\IEEEmembership{Fellow, ~IEEE}
    
    % <-this % stops a space
\thanks{This work was supported by Institute of Information \& communications Technology Planning \& Evaluation (IITP) grant funded by the Korea government (MSIT) (No. 2019-0-00079, Artificial Intelligence Graduate School Program (Korea University) and No. 2021-0-02068, Artificial Intelligence Innovation Hub) and Voice\&Avatar, Naver Cloud, Seongnam, Korea. \textit{(Corresponding author: Seong-Whan Lee.)}}
\thanks{Hyung-Seok Oh, Sang-Hoon Lee, and Seong-Whan Lee are with the Department of Artificial Intelligence, Korea University, Seoul 02841, South Korea (e-mail: hs\_oh@korea.ac.kr;sh\_lee@korea.ac.kr;sw.lee@korea.ac.kr).}

% \thanks{Manuscript received - -, 2023; }
}

% The paper headers
% \markboth{IEEE/ACM TRANSACTIONS ON AUDIO, SPEECH, AND LANGUAGE PROCESSING} % , Vol.-, 2023
% {Shell \MakeLowercase{\textit{et al.}}: A Sample Article Using IEEEtran.cls for IEEE Journals}

% \IEEEpubid{0000--0000/00\$00.00~\copyright~2023 IEEE}
% Remember, if you use this you must call \IEEEpubidadjcol in the second
% column for its text to clear the IEEEpubid mark.

\maketitle
\begin{abstract}
Expressive text-to-speech systems have undergone significant advancements owing to prosody modeling, but conventional methods can still be improved. 
Traditional approaches have relied on the autoregressive method to predict the quantized prosody vector; however, it suffers from the issues of long-term dependency and slow inference. 
This study proposes a novel approach called DiffProsody in which expressive speech is synthesized using a diffusion-based latent prosody generator and prosody conditional adversarial training. Our findings confirm the effectiveness of our prosody generator in generating a prosody vector. Furthermore, our prosody conditional discriminator significantly improves the quality of the generated speech by accurately emulating prosody. We use denoising diffusion generative adversarial networks to improve the prosody generation speed. Consequently, DiffProsody is capable of generating prosody 16 times faster than the conventional diffusion model. The superior performance of our proposed method has been demonstrated via experiments.
\end{abstract}

\begin{IEEEkeywords}
text-to-speech, speech synthesis, denoising diffusion model, prosody modeling, generative adversarial networks
\end{IEEEkeywords}

\section{Introduction}
\IEEEPARstart{R}{ecent} advancements in neural text-to-speech (TTS) models have significantly enhanced the naturalness of synthetic speech. 
In several studies\cite{8461368, NEURIPS2019_f63f65b5, ren2021fastspeech, wang17n_interspeech, NEURIPS2021_748d6b6e, Liu_Li_Ren_Chen_Zhao_2022, 9729483}, prosody modeling has been leveraged to synthesize speech that closely resembles human expression. Prosody, which encompasses  various speech properties, such as pitch, energy, and duration, plays a crucial role in the synthesis of expressive speech. 

In some studies\cite{pmlr-v80-skerry-ryan18a, 9414102}, reference encoders have been used to extract prosody vectors for prosody modeling. 
A global style token (GST)\cite{pmlr-v80-wang18h} is an unsupervised style modeling method that uses learnable tokens to model and control various styles. 
Meta-StyleSpeech\cite{pmlr-v139-min21b} proposes the application of style vectors extracted using a reference encoder through a style-adaptive layer norm.  
Progressive variational autoencoder TTS\cite{9747388} presents a method for gradual style adaptation. 
A zero-shot method for speech synthesis that comprises the use of a normalization architecture, speaker encoder, and feedforward transformer-based architecture\cite{9763046} was proposed. 
Despite the intuitive and effective nature of using a reference encoder, these methods cannot reflect the details of prosody in their synthetic speech without ground-truth (GT) prosody information. 

Recently, methods for inferring prosody from text in the absence of reference audio have been developed \cite{seshadri22_interspeech, bae22_interspeech}. 
FastPitch\cite{9413889}, for instance, synthesizes speech under text and fundamental frequency conditions. 
FastSpeech 2\cite{ren2021fastspeech} aims to generate natural, human-like speech by using extracted prosody features, such as pitch, energy, and duration, through an external tool and introduce a variance adaptor module that predicts these features.
Some studies\cite{9746253, chen2021adaspeech} have proposed hierarchical models through the design of prosody features at both coarse and fine-grained levels. 
However, the separate modeling of prosodic features may yield unnatural results owing to their inherent correlation. 

Some studies have predicted a unified prosody vector, thus enhancing the representation of prosody, given the interdependence of prosody features. 
Text-predicted GST\cite{8639682} is a method for modeling prosody without reference audio by predicting the weight of the style token from the input text.
\cite{9868130} proposed a method for paragraph-based prosody modeling by introducing a paragraph encoder. 
Gaussian-mixture-model-based phone-level prosody modelling\cite{9640518} is a method for sampling reference prosody from Gaussian components.
\cite{9420276} proposed a method for modeling prosody using style, perception, and frame-level reconstruction loss.
There are also studies in which prosody is modeled using pre-trained language models\cite{stephenson22_interspeech, yoon22b_interspeech, makarov22_interspeech, hwang2023pausespeech}.
ProsoSpeech \cite{9746883} models prosody with vector quantization (VQ) using large amounts of data and predicts the index of the codebook using an autoregressive (AR) prosody predictor. 
However, when predicting prosody vectors, an AR prosody predictor encounters challenges related to long-term dependencies.
\IEEEpubidadjcol

To address these issues, we propose DiffProsody, a novel approach that generates expressive speech by employing a diffusion-based latent prosody generator (DLPG) and prosody conditional adversarial training. 

The primary contributions of this work are as follows:
\begin{itemize}
    \item We propose a diffusion-based latent prosody modeling method that can generate high-quality latent prosody representations, thereby enhancing the expressiveness of synthetic speech. Furthermore, we adopted denoising diffusion generative adversarial networks (DDGANs) to reduce the number of timesteps, resulting in speeds that were 2.48× and 16× faster than those of the AR model and denoising diffusion probabilistic model (DDPM) \cite{NEURIPS2020_4c5bcfec}, respectively.
    \item We propose prosody conditional adversarial training to ensure an accurate reflection of prosody using the TTS module. A significant improvement in smoothness, attributable to vector quantization, was observed in the generated speech. 
    \item Objective and subjective evaluations demonstrated that the proposed method outperforms comparative models. 
\end{itemize}
The implementation\footnote{\url{https://github.com/hsoh0306/DiffProsody}} of proposed method and audio samples\footnote{\url{https://prml-lab-speech-team.github.io/demo/DiffProsody/}} for various datasets, such as VCTK\footnote{\url{https://datashare.ed.ac.uk/handle/10283/2651}} and LibriTTS\cite{zen19_interspeech}, are available online.

\section{Related works}
\subsection{Non-autoregressive text-to-speech}
Traditional TTS models function autoregressively. This implies that the spectrogram generates one frame at a time, with each frame conditioned on the preceding frames. Despite the high-quality speech that this approach can produce, it has a drawback in terms of speed owing to the sequential nature of the generation process.
To address this issue, non-autoregressive TTS (NAR-TTS) models have been proposed as an alternative for parallel generation. 
These models have the advantage of simultaneously generating the entire spectrogram, thus resulting in a significant acceleration of the speech synthesis. 
FastSpeech\cite{NEURIPS2019_f63f65b5} and FastSpeech 2\cite{ren2021fastspeech} serve as examples of NAR-TTS models that can synthesize speech at a much faster rate than their AR counterparts while maintaining a comparable level of quality.
For parallel generation, these models require phoneme-level durations. FastSpeech uses an AR teacher model to obtain durations through knowledge distillation. 
The phoneme-level input is scaled to the frame-level using a length regulator, and a transformer-based network \cite{NIPS2017_3f5ee243} is used to generate the entire utterance at once. 
FastSpeech 2 addresses some of the disadvantages of FastSpeech by extracting the phoneme duration from forced alignment as the training target instead of relying on the attention map of the AR teacher model and introducing more variation information in speech as conditional inputs. 
In contrast to these models that use an external aligner,\cite{NEURIPS2020_5c3b99e8, pmlr-v139-popov21a, kim2021conditional, NEURIPS2022_69c754f5} are parallel TTS models that use an internal aligner to model duration. 
These parallel-generation models exhibit faster and more robust generation than the AR models. 
In this study, we adopted a transformer-based NAR model with a simple structure to focus on prosody modeling. 

\subsection{Generative adversarial networks}
% GAN 설명 
Generative adversarial networks (GANs)\cite{NIPS2014_5ca3e9b1} are generative models in which generative and discriminative networks compete against each other. 
The objective of the generative network is to create samples that closely resemble the true data distribution, whereas the discriminative network strives to differentiate between the data sampled from the true and generated distributions. 

These two networks play a minimax game with the following value function $V(D, G)$:

\begin{multline}
\min_{G}\max_{D}V(D,G)=\mathbb{E}_{\mathbf{x}\sim p_{data}(\mathbf{x})}[\log{(D(\mathbf{x}))}]\\
+\mathbb{E}_{\mathbf{z}\sim p_{\mathbf{z}}(\mathbf{z})}[\log{(1-D(G(\mathbf{z})))}],
\end{multline}

where $G$ represents the generative network, and $D$ represents the discriminative network. 
The training process involves minimizing $D(1-G(z))$ to recognize $D$ as real for $G$, and maximizing $log(D(x))$ for $D$ to learn the likelihood of real data.  
The conditional generation model was obtained by introducing condition $c$ into both $G$ and $D$. In this study, we incorporated adversarial training conditions into prosody features to generate expressive speech.

\subsection{Denoising diffusion models}
% Diffusion 모델 설명
The denoising diffusion model is a generative model that gradually collapses data into noise and generates data from noise. The processes of collapsing and denoising data are called the forward and reverse processes, respectively. The forward process gradually collapses data $\mathbf{x}_0$ into noise over the $T$-step, with a predefined variance schedule $\beta_{t}$.

\begin{equation}
q(\mathbf{x}_{1:T}|\mathbf{x}_0)=\prod_{t=1}^{T} q(\mathbf{x}_t|\mathbf{x}_{t-1}),
\end{equation} 
where $q(\mathbf{x}_{t}|\mathbf{x}_{t-1}) := \mathcal{N}(\mathbf{x}_{t};\sqrt{1-\beta_t}\mathbf{x}_{t-1},\beta_t I)$. The reverse process is defined as follows:
\begin{equation}
p_{\theta}(\mathbf{x}_{0:T})=p(\mathbf{x}_T)\prod_{t=1}^{T}p_{\theta}(\mathbf{x}_{t-1}|\mathbf{x}_{t}).
\end{equation}

The reverse process is driven by the $\theta$ denoising model parameterized by a. The denoising model was optimized for a variational bound on the negative log-likelihood.

\begin{multline}
\mathbb{E}[-\log{p_\theta(\mathbf{x}_0)}]\leq \mathbb{E}_q[-\log \frac{p_\theta(\mathbf{x}_{0:T})}{q(\mathbf{x}_{1:T}|x_0)}]\\
=\mathbb{E}_q[-\log{p(\mathbf{x}_T)}-\sum_{t\geq1}\log{\frac{p_\theta(\mathbf{x}_{t-1}|\mathbf{x}_t)}{q(\mathbf{x}_t|\mathbf{x}_{t-1})}}] := L.
\end{multline}

\begin{figure*}[!t] % \vspace{-0.8cm}
  \centering
  \includegraphics[width=1.0\textwidth]{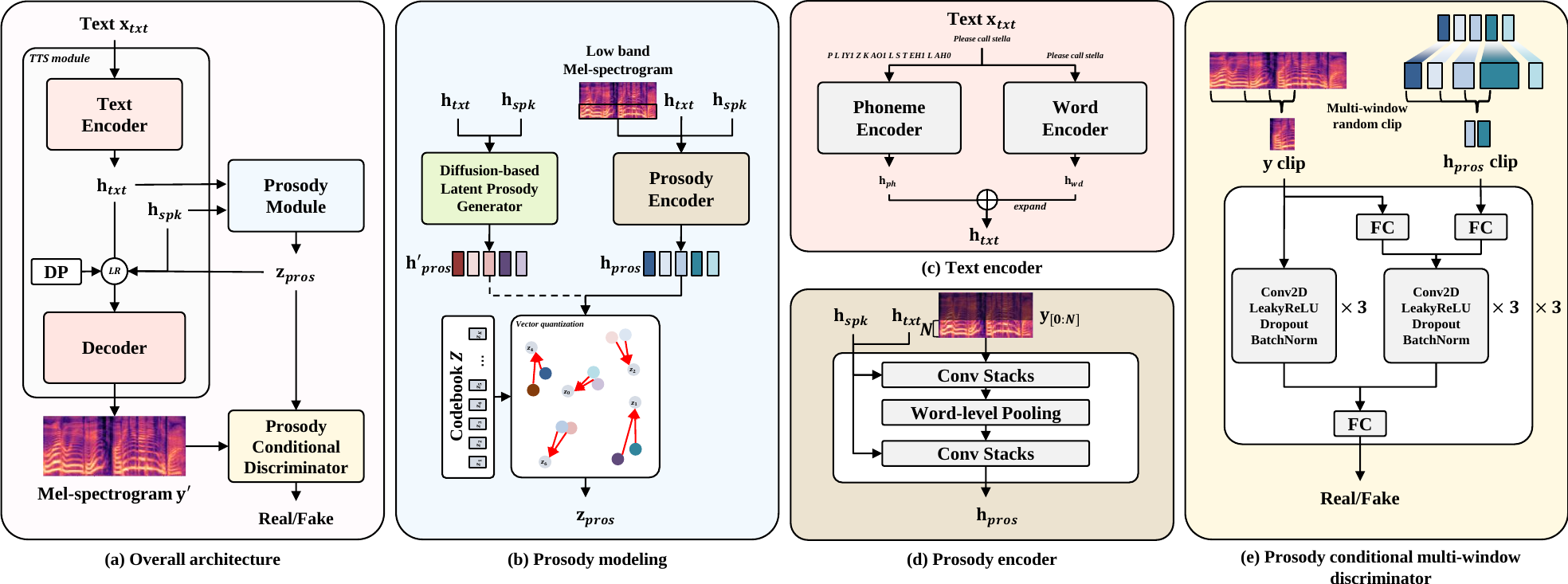}
  \caption{Framework of DiffProsody. (a) Overall architecture including TTS and prosody modeling with prosody conditional adversarial training; (b) Prosody modeling by vector quantization with prosody encoder and diffusion-based latent prosody generator; (c) Text encoder that models text at the phoneme-level and word-level; (d) Prosody encoder that models the word-level target prosody; (e) Prosody conditional discriminator for adversarial training. DP represents a duration predictor, and LR represents a length regulator. In the first stage, the TTS and prosody encoder are trained jointly, and in the second stage, a diffusion-based latent prosody generator (DLPG) is trained using the output of the pre-trained prosody encoder as a target. In inference, the TTS module synthesizes speech conditioned on the prosody vector generated by DLPG. } % \vspace{-0.65cm}
\label{model}
\end{figure*}

In the DDPM\cite{NEURIPS2020_4c5bcfec}, the denoising distribution $p_{\theta}(\mathbf{x}_{t-1}|\mathbf{x}_{t})$ is assumed to comprise a Gaussian distribution. 
Moreover, it has been demonstrated that the diffusion model can generate a diverse range of complex distributions, provided that a sufficient number of iterations are performed.
Proceeding with only a small step at a time is possible by setting the denoising distribution to a Gaussian distribution, which implies that a considerable number of timesteps are required.
DDGAN\cite{xiao2022tackling} was proposed by modeling the denoising distribution with a non-Gaussian multimodal distribution to reduce the sampling step. 
It predicts $\mathbf{x}_0$ with the generator $G_\theta$, which models the implicit distribution, in contrast to the denoising network of the DDPM, which predicts noise.
In the DDGAN, the conditional probability $p_\theta(\mathbf{x}_{t-1}|\mathbf{x}_t)$ is defined as follows:

\begin{multline}
p_\theta(\mathbf{x}_{t-1}|\mathbf{x}_t) :=\int{p_\theta(\mathbf{x}_{0}|\mathbf{x}_t)q(\mathbf{x}_{t-1}|\mathbf{x}_t, \mathbf{x}_0)}d\mathbf{x}_0 \\ =\int{p(\mathbf{z})q(\mathbf{x}_{t-1}|\mathbf{x}_t, \mathbf{x}_0=G_\theta(\mathbf{x}_t,\mathbf{z},t))}d\mathbf{z},
\end{multline}

where $p_\theta(\mathbf{x}_{t-1}|\mathbf{x}_t)$ represents the implicit distribution imposed by the generator $G_\theta(\mathbf{x}, \mathbf{z}, t)$, which outputs $\mathbf{x}_0$ given $\mathbf{x}_t$
and the latent variable $\mathbf{z}\sim p(\mathbf{z}):=\mathcal{N}(\mathbf{z};0,\mathbf{I})$.
In the DDGAN, the denoising distribution $p_\theta(\mathbf{x}_{t-1}|\mathbf{x}_t)$ is modeled as a complex multimodal distribution, in contrast to the unimodal distribution in the DDPM. 
Sampling $\mathbf{x}_0$ with small timesteps is possible by leveraging the complicated $p_\theta(\mathbf{x}_{t-1}|\mathbf{x}_t)$. 
We introduced this model to reduce the sampling timesteps  while maintaining its ability to generate diffusion models. 

\subsection{Prosody modeling}
Although the pronunciation capabilities of TTS models have seen significant advancements, they still fail to replicate the naturalness inherent in human speech.
Various studies have proposed methods for prosody modeling to address this limitation. 
One such method is the reference-based approach\cite{pmlr-v80-skerry-ryan18a, 9414102, pmlr-v80-wang18h,pmlr-v139-min21b, 9693198}, which is a type of expressive speech synthesis that extracts styles from the reference audio. 
This approach is particularly beneficial for style transfers, but can result in unnatural results when the text and reference audio do not align well.
However, there are methods that directly model prosody properties such as pitch, energy, and duration\cite{ren2021fastspeech, 9413889, lee2023pits, ju22_interspeech}. 
These methods offer the advantages of explainability and controllability because they directly use and model prosody features.
In the context of ProsoSpeech\cite{9746883}, the authors argue that prosodic features are interdependent and that modeling them separately can result in unnatural outcomes. 
To address this, they proposed the modeling of prosody as a latent prosody vector (LPV) and introduced an AR prosody predictor to obtain the LPV. 
In this study, we adopt a similar latent vector approach to model prosody.
The use of pre-trained language models was also suggested\cite{stephenson22_interspeech, yoon22b_interspeech, makarov22_interspeech, hwang2023pausespeech} . 
These models comprise the use of models that have been pre-trained on large datasets, such as BERT\cite{kenton2019bert} or GPT-3\cite{NEURIPS2020_1457c0d6}. 

\section{DiffProsody}
The proposed method, called DiffProsody, aims to enhance speech synthesis by incorporating a diffusion-based latent prosody generator (DLPG) and prosody conditional adversarial training. The overall structure and process of DiffProsody are presented in Figure \ref{model}.
In the first stage, we trained a TTS module and a prosody encoder using a text sequence and a reference Mel-spectrogram as inputs. 
The prosody conditional discriminator evaluates the prosody vector from the prosody encoder and the Mel-spectrogram from the TTS module to provide feedback on their quality.
In the second stage, we train a DLPG to sample a prosody vector that corresponds to the input text and speaker. 
During inference, the TTS module synthesizes speech without relying on a reference Mel-spectrogram. Instead, it uses the output of a DLPG. This facilitates the generation of expressive speech that accurately reflects the desired prosody.

\begin{figure*}[!t] % \vspace{-0.8cm}
  \centering
  \includegraphics[width=0.9\textwidth]{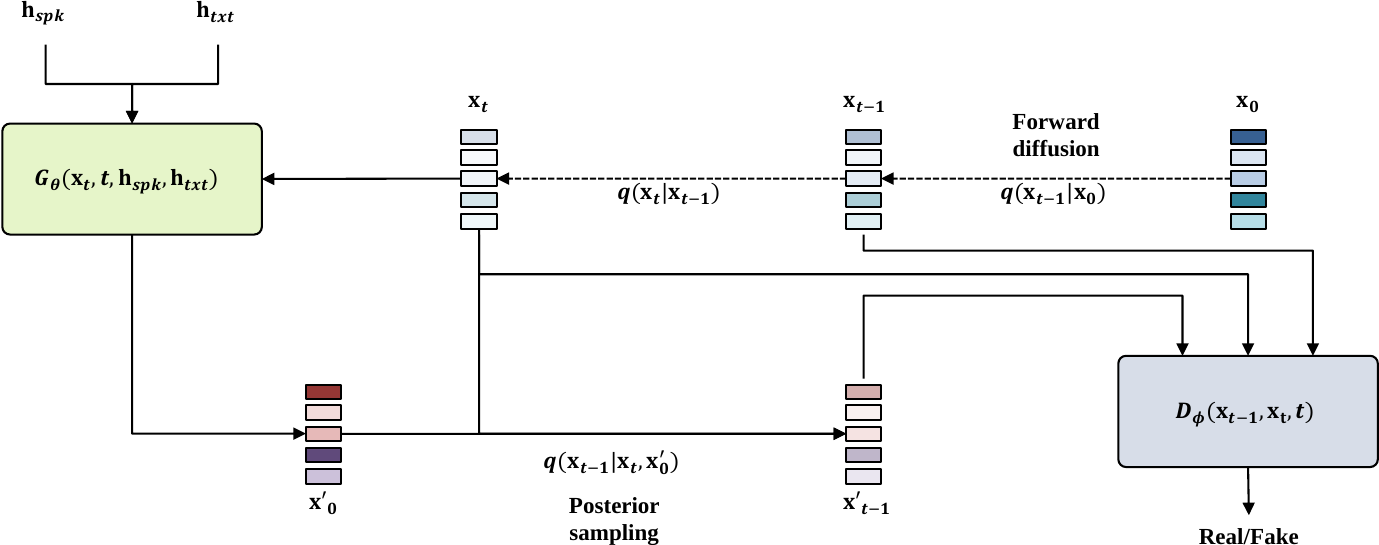}
  \caption{Training a diffusion-based latent prosody generator. We adopt the design of DDGANs\cite{xiao2022tackling} to shorten the diffusion timestep. The generator $G_\theta$ takes speaker hidden representation $\mathbf{h}_{spk}$ and text hidden representation$\mathbf{h}_{txt}$, timestep $t$, and noisy data $\mathbf{x}_{t}$ as input to generate $\mathbf{x}_{0}'$, and the disriminator $D_\phi$ determines which of $\mathbf{x}_{t-1}'$ obtained by posterior sampling on $\mathbf{x}_{0}'$ and $\mathbf{x}_{t-1}$ obtained by forward process on $\mathbf{x}_{0}$ is compatible with $\mathbf{x}_{t}$ at $t$ timestep. 
} % \vspace{-0.65cm}
\label{model2}
\end{figure*}

\subsection{Text-to-speech module}
The TTS module is designed to transform text into Mel-spectrograms using speaker and prosody vectors as conditions. The overall structure of the model is presented in Figure \ref{model}a. The TTS module comprises a text encoder and a decoder. The text encoder processes the text at both the phoneme and word levels, as illustrated in Figure \ref{model}c. The input text, denoted as $\mathbf{x}_{txt}$, is converted into a text hidden representation $\mathbf{h}_{txt}$, by the phoneme encoder $E_{p}$ and word encoder $E_{w}$. The $E_{p}$ takes the phoneme-level text $\mathbf{x}_{ph}$ and the $E_{w}$ takes as input the word-level text $\mathbf{x}_{wd}$. The $\mathbf{h}_{txt}$ is then obtained as the element-wise sum of the outputs of $E_{p}(\mathbf{x}_{ph})$ and $E_{w}(\mathbf{x}_{wd})$ expanded to the phoneme-level.

\begin{equation}
\mathbf{h}_{txt}=E_{p}(\mathbf{x}_{ph})+expand(E_{w}(\mathbf{x}_{wd})),
\end{equation}

where $expand$ is an operation that expands the word-level features to the phoneme-level.
Obtaining the quantized prosody vector $\mathbf{z}_{pros}$ involves using $\mathbf{h}_{txt}$ and speaker hidden representation $\mathbf{h}_{spk}$ as inputs for the prosody module. In addition, $\mathbf{h}_{spk}$ is acquired using a pre-trained speaker encoder. 
We use Resemblyzer\footnote{\url{https://github.com/resemble-ai/Resemblyzer}}, an open-source model trained with generalized end-to-end loss (GE2E)\cite{8462665}, to extract $\mathbf{h}_{spk}$. 

During the first stage of training, a prosody encoder is employed, which receives the target Mel-spectrogram. 
In the inference, $\mathbf{z}_{pros}'$ is obtained by inputting $\mathbf{h}_{txt}$ and $\mathbf{h}_{spk}$ into a DLPG, and this is performed without a reference Mel-spectrogram. Finally, the information related to the text, speaker, and prosody is combined by expanding the latent vectors $\mathbf{h}_{txt}$, $\mathbf{h}_{spk}$, and $\mathbf{z}_{pros}$ to the phoneme-level and then performing an element-wise summation. 

\begin{equation}
\mathbf{h}_{total}=\mathbf{h}_{txt} + \mathbf{h}_{spk} + \mathbf{z}_{pros}.
\end{equation}

The phoneme duration is modeled using the duration predictor $DP$. 
The goal of the $DP$ is to predict the phoneme duration at the frame-level based on the input variable $\mathbf{h}_{total}$.

\begin{equation}
dur'=DP(\mathbf{h}_{total}).
\end{equation}

In addition, there is a length regulator $LR$ that expands the input variable to the frame-level using the phoneme duration $dur$. The expanded $\mathbf{h}_{total}$ is then transformed to Mel-spectrogram $\mathbf{y}'$ by $D_{mel}$.

\begin{equation}
\mathbf{y}'=D_{mel}(LR(\mathbf{h}_{total}, dur)).
\end{equation}

For TTS modeling, we use two types of losses: the mean square error (MSE) and structural similarity index (SSIM) loss. These losses aid in accurately modeling the TTS. For the duration modeling, we use the MSE loss.

\begin{equation}
\mathcal{L}_{rec}=\mathcal{L}_{MSE}(\mathbf{y}, \mathbf{y}') + \mathcal{L}_{SSIM}(\mathbf{y}, \mathbf{y}').
\end{equation}

\begin{equation}
\mathcal{L}_{dur}=\mathcal{L}_{MSE}(dur, dur').
\end{equation}

\subsection{Prosody module}
Figure \ref{model}b presents the prosody module, which includes a prosody encoder $E_{pros}$ that derives a prosody vector from a reference Mel-spectrogram, a DLPG that produces a prosody vector using text and speaker hidden states, and a codebook $Z=\{\mathbf{z}_k\}_{k=1}^{K}\in \mathbb{R}^{K\times d_{z}}$, where $K$ represents the size of the codebook and $d_{z}$ is the dimension of the codes.
During the training of $E_{pros}$, instead of a full-band Mel-spectrogram, we used a low-frequency band Mel-spectrogram to alleviate disentanglement, as in the case of ProsoSpeech\cite{9746883}. Figure \ref{model}d presents the structure of $E_{pros}$, which comprises two convolutional stacks and a word-level pooling layer. 
To extract the target prosody, $E_{pros}$ uses the lowest $N$ bins of the target Mel-spectrogram $\mathbf{y}_{[0:N]}$, along with the $\mathbf{h}_{txt}$ and $\mathbf{h}_{spk}$, as its inputs. The output of this process is a prosody vector, $\mathbf{h}_{pros} \in \mathbb{R}^{L\times d_{z}}$, where $L$ is the word-level length of the input text.

\begin{equation}
\mathbf{h}_{pros}=E_{pros}(\mathbf{y}_{[0:N]}, \mathbf{h}_{txt}, \mathbf{h}_{spk}).
\end{equation}

During the inference stage, the prosody vector $\mathbf{h}_{pros}'$ is obtained using the prosody generator trained in the second stage.

\begin{equation}
\mathbf{h}_{pros}'=DLPG(\mathbf{h}_{txt}, \mathbf{h}_{spk}).
\end{equation}

The $DLPG$ process is described in section \ref{dlpg}.
To obtain the discrete prosody token sequence $\mathbf{z}_{pros} \in \mathbb{R}^{L\times d_{z}}$, the vector quantization layer $Z$ maps each prosody vector $\mathbf{h}_{pros}^{i}\in \mathbb{R}^{d_{z}}$ to the nearest element of the codebook entry $\mathbf{z}_{k} \in \mathbb{R}^{d_{z}}$.

% \begin{equation}
% \mathbf{z}_{pros}=\mathbf{z}_k \text{ where } k = argmin_{j}||\mathbf{h}_{pros} - \mathbf{z}_{j}||_2
% \end{equation}
\begin{equation}
\mathbf{z}_{pros}^{i} = \argmin_{\mathbf{z}_{k}\in Z}||\mathbf{h}_{pros}^{i} - \mathbf{z}_{k}||_2 \text{ for } i = 1 \text{ to } L,
\end{equation}
where $\mathbf{z}_{pros}^i$ is $i$-th element of $\mathbf{z}_{pros}$.
In the first stage, the TTS module is trained jointly with the codebook $Z$ and prosody encoder $E_{pros}$.

\begin{equation}
\mathcal{L}_{vq}=||sg[\mathbf{h}_{pros}]-\mathbf{z}_{pros}||_2^2+\beta||\mathbf{h}_{pros}-sg[\mathbf{z}_{pros}]||_2^2,
\end{equation}

where $sg[\cdot]$ denotes the stop-gradient operation. 
Moreover, we employ an exponential moving average (EMA)\cite{NIPS2017_7a98af17} to enhance the learning efficiency by applying it to codebook updates.

\subsection{Prosody conditional adversarial training}
For our prosody adversarial training, we develop prosody conditional discriminators (PCDs) that handle inputs of varying lengths. This design is inspired by multi-length window discriminators\cite{ren2022revisiting}. The PCD structure is presented in Figure \ref{model}e.
The PCD is designed to accept a Mel-spectrogram $y$ and a quantized prosody vector $\mathbf{z}_{pros}$ as inputs, and its role is to determine whether these input features are original or generated. The PCD comprises two lightweight convolutional neural networks (CNNs) and fully connected layers. One of the CNNs is designed to receive only the Mel-spectrogram, whereas the other is designed to receive a combination of $\mathbf{z}_{pros}$ and $y$.
To match the corresponding PCD, the length of each Mel-spectrogram and the extended $\mathbf{z}_{pros}$ are randomly clipped. For our objective function, we adopt the least square GAN loss\cite{Mao_2017_ICCV}:

\begin{multline}
\mathcal{L}_{D}=\sum_{i} [\mathbb{E} [(PCD^i(\mathbf{y}', \mathbf{z}_{pros}))^2]\\
+ \mathbb{E} [(PCD^i(\mathbf{y}, \mathbf{z}_{pros}) - 1)^2]],
\end{multline}
\begin{equation}
\mathcal{L}_{G}=\sum_{i} \mathbb{E} [(PCD^i(\mathbf{y}', \mathbf{z}_{pros}) - 1)^2],
\end{equation}

where $\mathcal{L}_{D}$ denotes the training goal of the discriminators and $\mathcal{L}_{G}$ represents the feedback on the TTS module. 
The final object $\mathcal{L}_{TTS}$ of the TTS module is as follows:

\begin{equation}
\mathcal{L}_{TTS}=\mathcal{L}_{rec} + \mathcal{L}_{dur} + \mathcal{L}_{vq} + \lambda_1 \mathcal{L}_{G},
\end{equation}

where $\lambda_1$ corresponds to the weight of the adversarial loss.

\subsection{Diffusion-based latent prosody generator} \label{dlpg}
We propose a new module called the DLPG, which leverages the powerful generative capabilities of diffusion models. 
In addition, we introduce a DDGAN framework\cite{xiao2022tackling}, which enables faster sampling by reducing the number of required timesteps. 
Figure \ref{model2} presents the training process for the DLPG. 
During training, the DLPG aims to generate target $\mathbf{h}_{pros}$, which is extracted from the prosody encoder trained in the first stage. The DLPG is trained to produce $\mathbf{h}_{pros}'$ based on the $\mathbf{h}_{spk}$ and $\mathbf{h}_{txt}$. 
In the diffusion model, we set $\mathbf{x}_0$ as the target $\mathbf{h}_{pros}$. 
The DLPG generator $G_\theta$ directly generates $\mathbf{x}'_0$.
 
\begin{equation}
\mathbf{x}'_0=G_\theta(\mathbf{x}_{t},t, \mathbf{h}_{spk},\mathbf{h}_{txt}),
\end{equation}

where $t$ is timestep of diffusion process.
To ensure adversarial training, $\mathbf{x}'_{t-1}$ is derived from $\mathbf{x}_{t}$ and $\mathbf{x}'_{0}$ using posterior sampling $q(\mathbf{x}'_{t-1}|\mathbf{x}_t,\mathbf{x}'_0)$. Subsequently, a time-dependent discriminator $D_\phi$ determines the compatibility of $\mathbf{x}_{t-1}$ (obtained from the forward processing of $\mathbf{x}_0$) and $\mathbf{x}'_{t-1}$ (generated through posterior sampling of $\mathbf{x}_{0}'$) with respect to $t$ and $\mathbf{x}_t$, conditioned on $\mathbf{h}_{spk}$ and $\mathbf{h}_{txt}$. The objective function of $G_{\theta}$ is then defined as follows:

\begin{equation}
\mathcal{L}_{G_\theta}^{adv}=\sum_{t\geq1}\mathbb{E} [(D_\phi(\mathbf{x}'_{t-1},\mathbf{x}_{t},t,\mathbf{h}_{txt}, \mathbf{h}_{spk}) - 1)^2],
\end{equation}

\begin{equation}
\mathcal{L}_{G_\theta}^{rec}=L_{MAE}(\mathbf{x}_0,\hat{\mathbf{x}}_0),
\end{equation}

where $\mathcal{L}_{G_\theta}^{adv}$ corresponds to the adversarial loss, and $\mathcal{L}_{G_\theta}^{rec}$ denotes the reconstruction loss of $G_\theta$. The total generator loss $\mathcal{L}_{G_\theta}$ is expressed as follows: 

\begin{equation}
\mathcal{L}_{G_\theta}=\mathcal{L}_{G_\theta}^{rec}+\lambda_2 \mathcal{L}_{G_\theta}^{adv},
\end{equation}

where $\lambda_2$ is the weight of adversarial loss $\mathcal{L}_{G_\theta adv}$. The objective function of the $D_{\phi}$ is as follows:

\begin{multline}
\mathcal{L}_{D_\phi}=\sum_{t\geq1}[\mathbb{E} [D_\phi(\mathbf{x}'_{t-1},\mathbf{x}_{t},t,\mathbf{h}_{txt}, \mathbf{h}_{spk})^2] \\ +\mathbb{E} [(D_\phi(\mathbf{x}_{t-1},\mathbf{x}_{t},t,\mathbf{h}_{txt}, \mathbf{h}_{spk})-1)^2]].
\end{multline}

The DLPG leverages the DDGAN framework to achieve stable and high-quality results in only a few timesteps. This process involves the $G_\theta$, which iteratively generates $\mathbf{x}'_{0}$ $T$ times during the inference. We set $\mathbf{x}_T$ to follow a normal distribution. 
% After the first iteration, $G_\theta$ receives the posterior sampled $\mathbf{x}'_{t-1}$ as an input from the generated $\mathbf{x}'_{0}$. 
The $\mathbf{h}_{pros}'$ is obtained as the final $\mathbf{x}'_{0}$ of the reverse process. 
The final prosody vector $\mathbf{z}_{pros}$ is then derived through the vector quantization of $\mathbf{h}_{pros}'$.

\subsection{Inference}
Here is the step-by-step process of generating a Mel-spectrogram using the trained TTS module and DLPG.
\begin{enumerate}
 \item The $\mathbf{h}_{txt}$ is extracted from the text encoder, and $\mathbf{h}_{spk}$ is extracted from the pre-trained speaker encoder.
 \item The DLPG generates a $\mathbf{h}_{pros}'$ with $\mathbf{h}_{txt}$ and $\mathbf{h}_{spk}$ as inputs.
 \item $\mathbf{h}_{pros}'$ is mapped to a codebook, denoted as $Z$, in the VQ layer to obtain the prosody vector, which is denoted as $\mathbf{z}_{pros}$.
 \item The decoder $D_{mel}$ generates a Mel-spectrogram $\mathbf{y}'$ using $\mathbf{h}_{txt}$, $\mathbf{h}_{spk}$ and $\mathbf{z}_{pros}$. This process involves the expansion $\mathbf{h}_{txt}$, $\mathbf{h}_{spk}$, and $\mathbf{z}_{pros}$ to the frame-level. The phoneme-duration is predicted by the duration predictor.
 \item The Mel-spectrogram $y'$ was converted to a raw waveform using a pre-trained vocoder. 
\end{enumerate}

\section{Experimental result and discussion}
\subsection{Experimental setup}
We conducted experiments using the VCTK dataset, a multispeaker English dataset consisting of audio clips of 44,200 sentences recorded by 109 speakers for approximately 400 sentences. A total of 2,180 audio clips were randomly selected with 20 sentences per speaker, which constituted the test set. Furthermore, 545 audio clips randomly selected from five sentences per speaker were used as the verification sets, and the remainder were used as the training sets. 
We sampled audio at 22,050 Hz and then transformed it to an 80-bin Mel-spectrogram using an STFT with a window length of 1,024 and hop size of 256. 
The text was converted into phoneme sequences for text input using a grapheme-to-phoneme tool\footnote{\url{https://github.com/Kyubyong/g2p}}. We extracted the phoneme duration using the Montreal Forced Aligner\cite{mcauliffe17_interspeech} tool. 
Furthermore, we used the AdamW optimizer\cite{loshchilov2018decoupled} with $\beta_{1}=0.9$ and $\beta_{2}=0.98$. The learning rates for the TTS and latent prosody generator training were set as $5\times 10^{-4}$ and $2\times 10^{-4}$, respectively. Throughout the training process, the batch size used was 48. The TTS and prosody encoder were updated in 160k steps, and the latent prosody generator was updated in 320k steps. 
In the experiment, all the audio was synthesized using the official implementation of the HiFi-GAN\footnote{\url{https://github.com/jik876/hifi-gan}}\cite{NEURIPS2020_c5d73680} and a pre-trained model. 
% Speaker embeddings were extracted using the pre-trained voice encoder of the Resemblyzer\footnote{\url{https://github.com/resemble-ai/Resemblyzer}}.
We trained the TTS module and prosody encoder for approximately 16 h and the DLPG for 7 h using a single NVIDIA RTX A6000 GPU.

\subsection{Implementation details}
The number of layers, hidden size, filter size, and kernel size of the feed-forward transformer blocks of the phoneme encoder, word encoder, and decoder were set as 4, 192, 384, and 5, respectively. The extracted speaker embedding was projected onto 192 dimensions, and the number of dimensions of the prosody vector was 192. 
The structure of the prosody encoder follows that of ProsoSpeech\cite{9746883}.
For VQ, we set the size of codebook and dimension of code as 128 and 192, respectively. and updated it using EMA with a decay rate of 0.998. The codebook was initialized as the center of the k-means clustering after 20k steps of TTS training. The number of Mel-spectrogram bins used in the prosody encoder $N$ was set as 20. 
We set multiple PCDs to receive different input sizes, such as\cite{Lee_Yoon_Noh_Kim_Lee_2021, ye2022syntaspeech, 9747155}. The PCD comprises two 2D convolution stacks and three fully connected layers. 
The convolution stacks in the PCD consist of three 2D convolutions: LeakyReLU, BatchNorm, and a linear layer with a multilength window size of [32, 64, 128]. The latent prosody generator consists of 20 residual blocks with a hidden size of 384. 
The prosody discriminator is configured with four convolution layers and hidden dimensions of 384.
We investigated $\lambda_{1}$ and $\lambda_{2}$ between 0.001 and 1.0 and set $\lambda_{1}$ and $\lambda_{2}$ as 0.01 and 0.05, respectively.
The number of timesteps used in the training and inference was set as four.

\subsection{Comparative studies}
We developed a prosody model from previous works, our proposed model, and the model of the ablation study for performing 
comparative experiments. All the models produced an 80-bin Mel-spectrogram and synthesized speech using the same vocoder.

\subsubsection{GT}
GT audio is human-recorded.
\subsubsection{GT (vocoded)}
The generated audio was obtained by converting the GT Mel-spectrogram using HiFi-GAN V1\cite{NEURIPS2020_c5d73680}.
\subsubsection{FastSpeech 2}
FastSpeech 2\cite{ren2021fastspeech} is a speech-synthesis model that directly predicts prosody features (pitch and energy). We trained FastSpeech 2 using an open-source implementation\footnote{\url{https://github.com/NATSpeech/NATSpeech}}. For the purpose of a fair comparison, we used a text encoder at the phoneme and word-levels.
\subsubsection{ProsoSpeech}
ProsoSpeech\cite{9746883} is a speech synthesis model that models prosody features as latent prosody vectors and predicts them using an AR predictor. We implemented the model by following the hyperparameters in the study and used the model that provided the best performance. 
\subsubsection{DiffProsody}
The proposed model was implemented using a text encoder and decoder with the same structure as FastSpeech 2 and ProsoSpeech. We used a prosody encoder with the same structure as that of ProsoSpeech. In the first stage, a prosody conditional discriminator was added. In the second stage, a DLPG was used instead of an AR predictor. 
\subsubsection{DiffProsody (AR)}
DiffProsody (AR) is a model in which a DiffProsody trained with an AR prosody predictor to estimate prosody vectors. It has the same structure as the ProsoSpeech prosody predictor.
\subsubsection{DiffProsody (DDPM)}
DiffProsody (DDPM) is a model in which a DiffProsody trained with a DDPM framework to estimate a prosody vector. It has the same structure as DiffProsody's diffusion-based latent generator.
\subsubsection{DiffProsody (w/o PCD)}
DiffProsody (w/o PCD) is a model in which a DiffProsody trained without the assistance of a prosody conditional discriminator.
\subsubsection{DiffProsody (w/o VQ)}
DiffProsody (w/o VQ) is a model in which a DiffProsody trained without a VQ layer in the prosody encoder. 
\subsubsection{DiffProsody N}
DiffProsody $N$ is a model in which a DiffProsody trained with the lowest $N$ Mel-bins in the prosody encoder.

\subsection{Subjective metrics}
A subjective assessment is conducted to confirm the effectiveness of the proposed method. To measure the level of naturalness, we used the mean opinion score (MOS). For this evaluation, we employed Amazon Mechanical Turk (MTurk), a crowdsourcing service, to gather feedback from 20 native Americans. MOS was assessed using a 5-point scale, and confidence intervals were calculated at a 95\% level. For the evaluation, 100 samples were randomly selected from the test set.

\subsection{Objective metrics}
For realizing an objective evaluation, we calculated the equal error rate (EER) using a pre-trained speaker verification model\footnote{\url{https://github.com/clovaai/voxceleb_trainer}}
\cite{jung22_interspeech}. We used the pre-trained wav2vec 2.0\cite{NEURIPS2020_92d1e1eb} to the compute character error rate (CER) and word error rate (WER). 

For the prosodic evaluation, we computed the average differences in utterance duration (DDUR)\cite{8607053, 8936924}, pitch error (RMSE$_{f_0}$) (in cents), periodicity error (RMSE$_{period}$) and F1 score of the voiced/unvoiced classification (F1$_{v/uv}$). We used torchcrepe\footnote{\url{https://github.com/maxrmorrison/torchcrepe}} to extract the pitch and periodicity features for evaluation.
In addition, we measured the Kullback–Leibler (KL) divergence of log f0 and log energy to compare the distributions of the prosody features in the generated audio. Finally, we calculated the real-time factor (RTF) to compare the generation speeds.

\subsubsection {Average differences in the utterance duration (DDUR)}
We obtained $DDUR$ by calculating the mean absolute error of the difference in the duration of each utterance. 

\begin{equation}
  DDUR=\frac{1}{N}\sum_{i=1}^N|dur_{i}-dur_{i}'|,
\end{equation}
where $dur_{i}$ denotes the duration of the $i$-th GT utterance and $dur_{i}'$ denotes the duration of the $i$-th generated utterance. 

\subsubsection {Pitch error} 
To measure the pitch error RMSE$_{f_0}$, we aligned the pitches extracted in hertz with dynamic time warping (DTW)\cite{muller2007dynamic} and calculated the root mean square (RMSE) in cents, defined as $1200\log_2(y/\hat{y})$ for pitch of the GT speech $y$ and pitch of the generated speech $\hat{y}$. We measured the portion wherein both the GT speech and generated speech were voiced.

\begin{equation}
  RMSE_{f_{0}}=\sqrt{\frac{1}{T}\sum_{i=1}^{T}(1200\log_2(y_i/y_{i}'))^2}.
\end{equation}

\subsubsection{Periodicity error}
We measured the periodicity error RMSE$_{period}$ by root mean square between the periodicities $\psi$ aligned with the DTW. 
% To measure the periodicity error RMSE$_{period}$, We calculated it by root mean square between periodicities $\psi$ aligned with the DTW. 

\begin{equation}
  RMSE_{period}=\sqrt{\frac{1}{T}\sum_{i=1}^{T}(\psi_i-\psi_{i}')^2},
\end{equation}

where $\psi_{i}$ means the $i$-th periodicity value.

\subsubsection{F1 score of voiced/unvoiced} 
We obtained voiced/unvoiced flags from the aligned pitches using DTW and calculated the F1 score (F1$_{v/uv}$) between them. We defined a match between the GT voiced flag $u$ and generated voiced flag $u'$ as a true positive (TP), a match between the GT unvoiced flag $uv$ and generated voiced flag $v'$ as a false positive (FP), and a match between the GT voiced flag $v$ and generated unvoiced flag $uv'$ as 
 a false negative (FN). 
 
\begin{equation}
  TP=\sum_{i}^{n}[v_{i} = v_{i}'], 
\end{equation}
\begin{equation}
  FP=\sum_{i}^{n}[uv_{i} = v_{i}'], 
\end{equation}
\begin{equation}
  FN=\sum_{i}^{n}[v_{i} = uv_{i}'],
\end{equation}

where $n$ is the length of the sequence, and $[a_i=b_i]$ is a function that returns 1 if the $i$-th element has the same value, and 0 otherwise. The precision and recall are defined as follows:

\begin{equation}
  precision=\frac{TP}{TP+FP}, recall=\frac{TP}{TP+FN}.
\end{equation}
We then calculated the F1 score as follows:
\begin{equation}
  F1_{v/uv}=\frac{2}{recall^{-1}+precision^{-1}}.
\end{equation}

\subsubsection{KL divergence of log f0 / log energy} 
We also measured the KL D to analyze the pitch and energy distribution. We first extracted the pitch and energy in log-scale. We then binned the entire range into 100 bins and applied a kernel density estimation\cite{terrell1992variable} to each bin to calculate the KL D for a smoothed distribution. The KL divergence for feature $x$ is defined as follows:

\begin{equation}
  KLD_{x}=\frac{1}{N}\sum_{i=1}^{N}(KLD(KDE(x_i), KDE(x_{i}')),
\end{equation}

where $N$ is the number of bins, $x_i$ is the probability in the $i$-th bin of the distribution of $x$, $KDE$ is the kernel density estimator function, and $KLD$ is the KL divergence calculation. 

\begin{table*}[!ht]
  \caption{Objective and subjective evaluation of previous methods} % \vspace{-0.3cm}
  \centering

  \resizebox{1.0\textwidth}{!}{
  \begin{tabular}{l|c|ccc|cccc|cc}
    \toprule
    \textbf{Method}           & \textbf{MOS (CI)} ($\uparrow$) & \textbf{CER (\%)} ($\downarrow$) & \textbf{WER (\%)} ($\downarrow$) & \textbf{DDUR} ($\downarrow$) & \textbf{EER (\%)} ($\downarrow$) & \textbf{RMSE$_{f0}$} ($\downarrow$) & \textbf{RMSE$_{period}$} ($\downarrow$) & \textbf{F1$_{v/uv}$} ($\uparrow$) & \textbf{RTF} ($\downarrow$) & \textbf{Params}\\ \midrule
      GT               & 4.24 ($\pm$ 0.02) & 0.63 & 1.53 & - & 2.184 & - & - & - & - & -\\
      GT (Vocoded)          & 4.17 ($\pm$ 0.03) & 0.76 & 1.74 & - & 2.539 & 22.80 & 0.336 & 0.7923 & - & - \\ 
      \midrule
      FastSpeech2           & 3.84 ($\pm$ 0.03) & 1.81 & 4.89 & 0.305 & 8.685 & 88.83 & 0.486 & 0.6848 & \textbf{0.015} & 27M\\
      ProsoSpeech           & 3.97 ($\pm$ 0.04) & 1.68 & 3.82 & 0.296 & 7.200 & 59.66 & 0.483 & 0.6874 & 0.149 & 59M \\
      DiffProsody           & \textbf{4.03} ($\pm$ 0.03) & \textbf{0.90} & \textbf{2.55} & \textbf{0.295} & \textbf{5.404} & \textbf{54.60} & \textbf{0.475} & \textbf{0.6952} & 0.054 & 53M \\ 
    \bottomrule
  \end{tabular}
   }
  \label{table:mos} % \vspace{-0.6cm}
\end{table*}

\subsection{Evaluation results}
Table \ref{table:mos} lists the MOS results and objective evaluations. The results demonstrate that the proposed model DiffProsody surpasses the other models in terms of both subjective and objective metrics. In particular, our model achieved a superior MOS for the subjective metrics, with a p-value of less than 0.05. It also exhibited outstanding performance in terms of the DDUR, EER, RMSE$_{f_0}$, RMSE$_{period}$, and F1$_{v/uv}$, which suggests that the speech generated by DiffProsody closely resembles the target prosody. Furthermore, our model displayed a lower WER and CER, thus indicating its capability for synthesizing speech with a more accurate pronunciation.

\begin{figure}[!t]
\centering
% \subfloat[]{\includegraphics[width=0.5\columnwidth]{figures/energy_hist_vocoded.png}%
% \label{fig_log_energy_vocoded}}
\captionsetup[subfloat]{labelfont=scriptsize,textfont=scriptsize}  % subfloat caption setting
\subfloat[Ground truth]{\includegraphics[width=0.9\columnwidth]{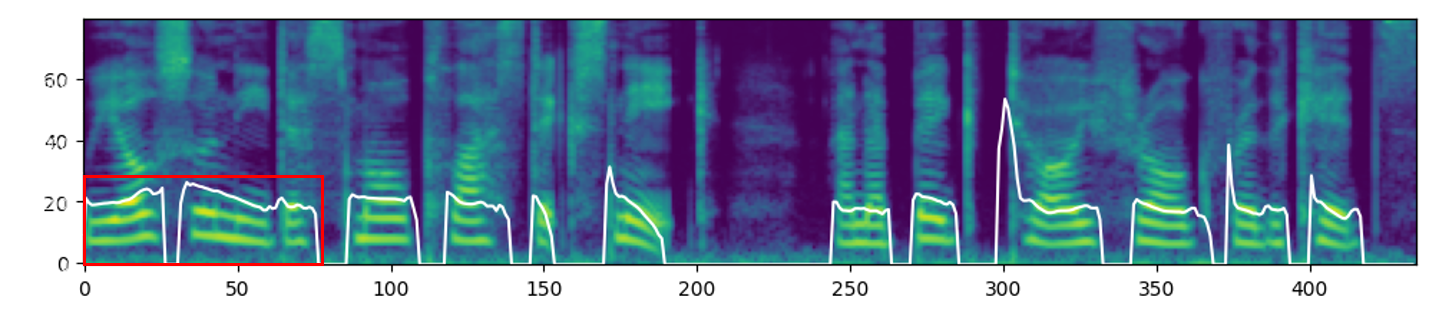}%
\label{fig_spec_gt}}

\subfloat[FastSpeech2]{\includegraphics[width=0.9\columnwidth]{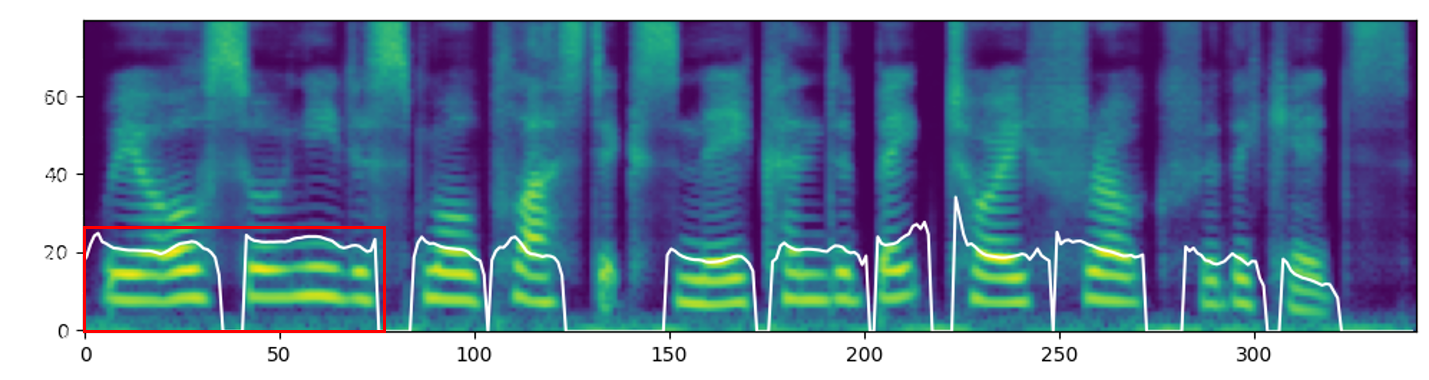}%
\label{fig_spec_fs2}}

\subfloat[ProsoSpeech]{\includegraphics[width=0.9\columnwidth]{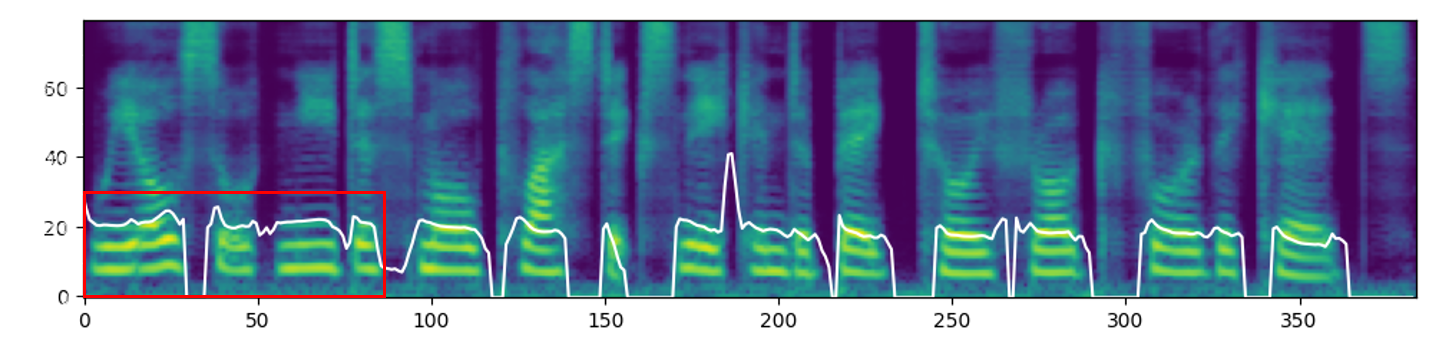}%
\label{fig_spec_prs}}

\subfloat[DiffProsody]{\includegraphics[width=0.9\columnwidth]{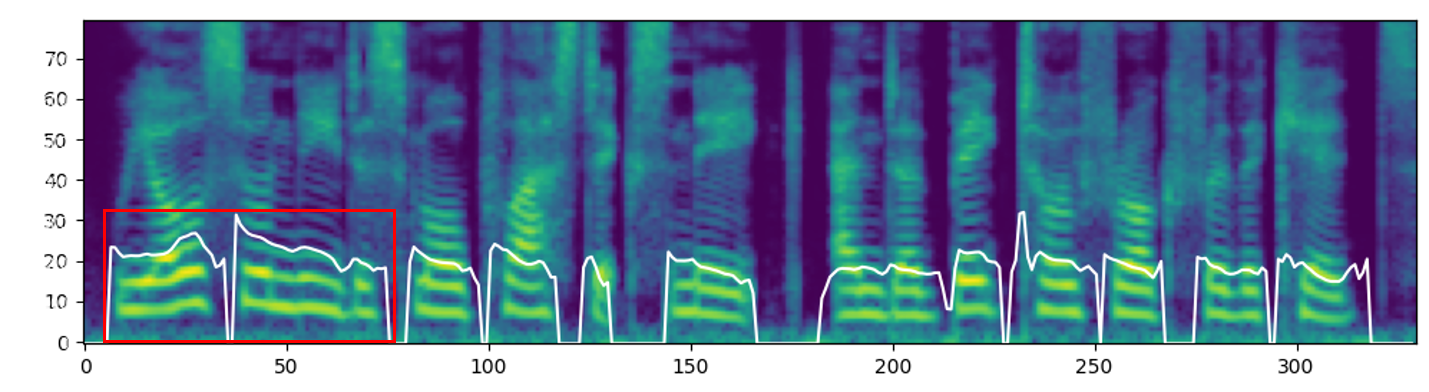}%
\label{fig_spec_dp}}
\caption{Comparison of the visualized spectrogram and pitch contour. The red box indicates that the proposed model is more similar to the GT.}
\label{fig_spec}
\end{figure}

Figure \ref{fig_spec} presents the Mel-spectrogram and pitch contour of speech from each model. The red box in the figure indicates that the DiffProsody model was more similar to the GT. To further evaluate the prosody of the generated speech, we examined pitch and energy distributions. 

\begin{figure}[!t]
\centering
\captionsetup[subfloat]{labelfont=scriptsize,textfont=scriptsize}  % subfloat caption setting
\subfloat[FastSpeech2]{\includegraphics[width=0.3\columnwidth]{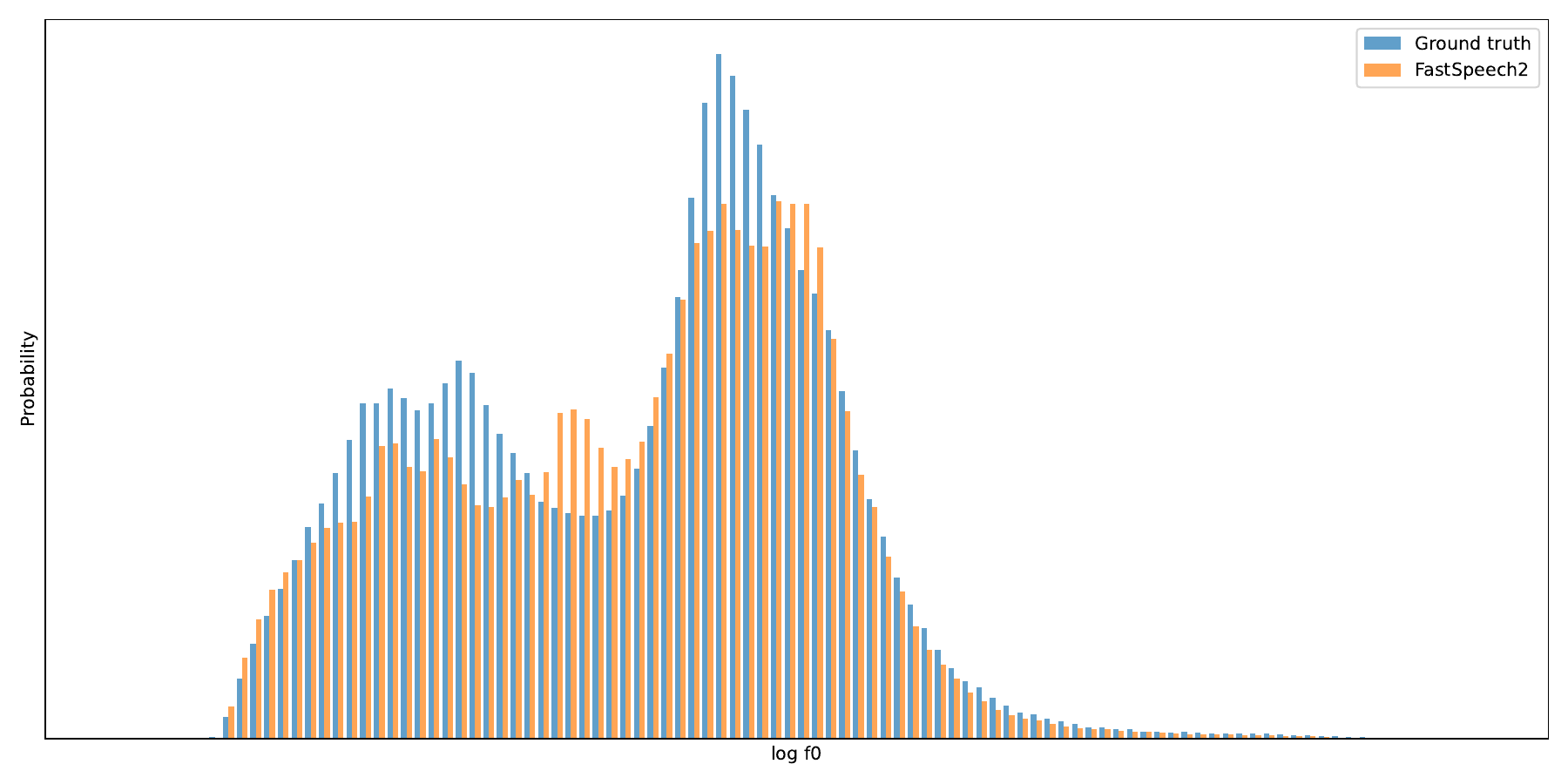}%
\label{fig_log_f0_fs2}}
\hspace{0.1cm}
\subfloat[ProsoSpeech]{\includegraphics[width=0.3\columnwidth]{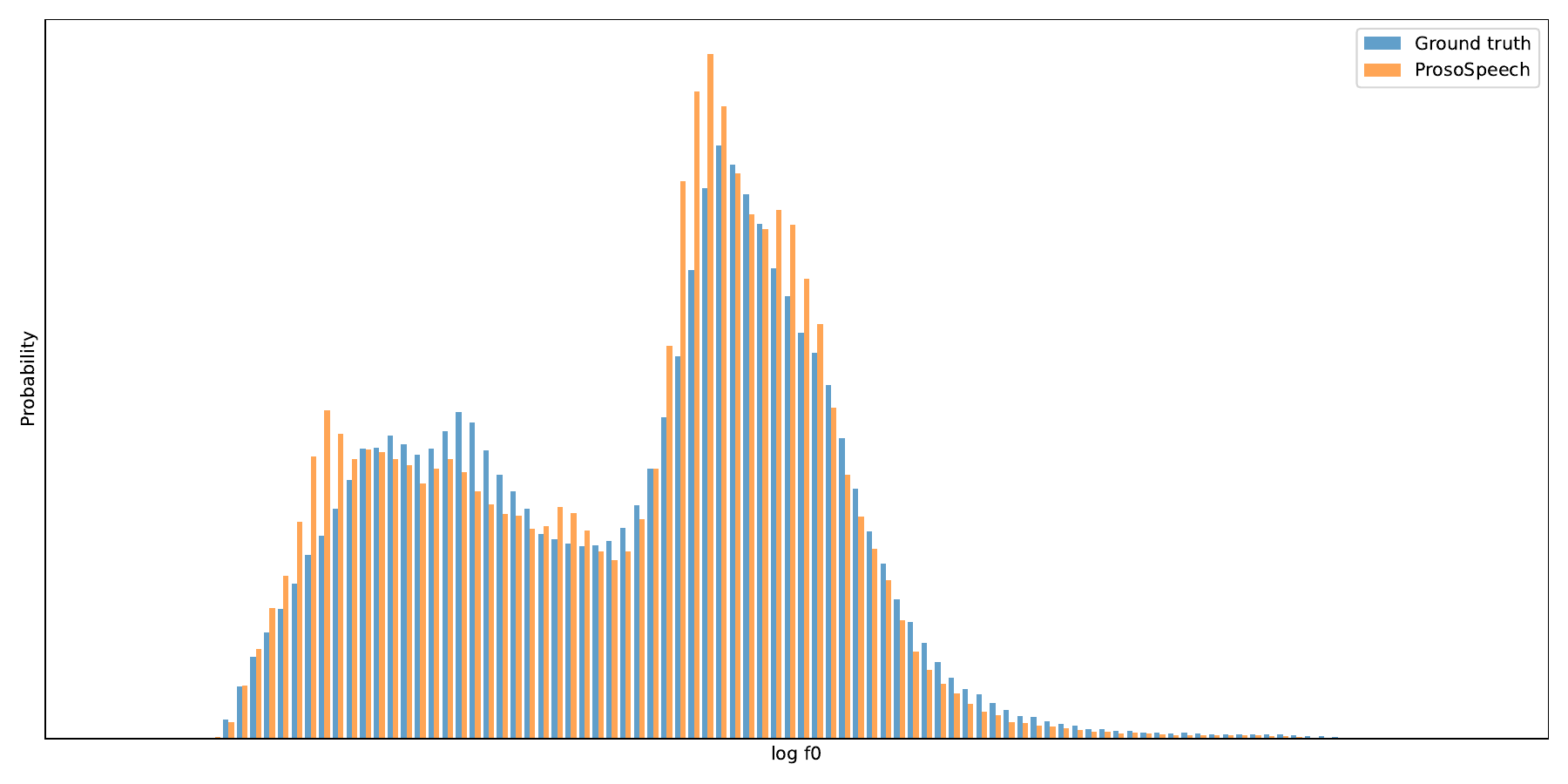}%
\label{fig_log_f0_prs}}
\hspace{0.1cm}
\subfloat[DiffProsody]{\includegraphics[width=0.3\columnwidth]{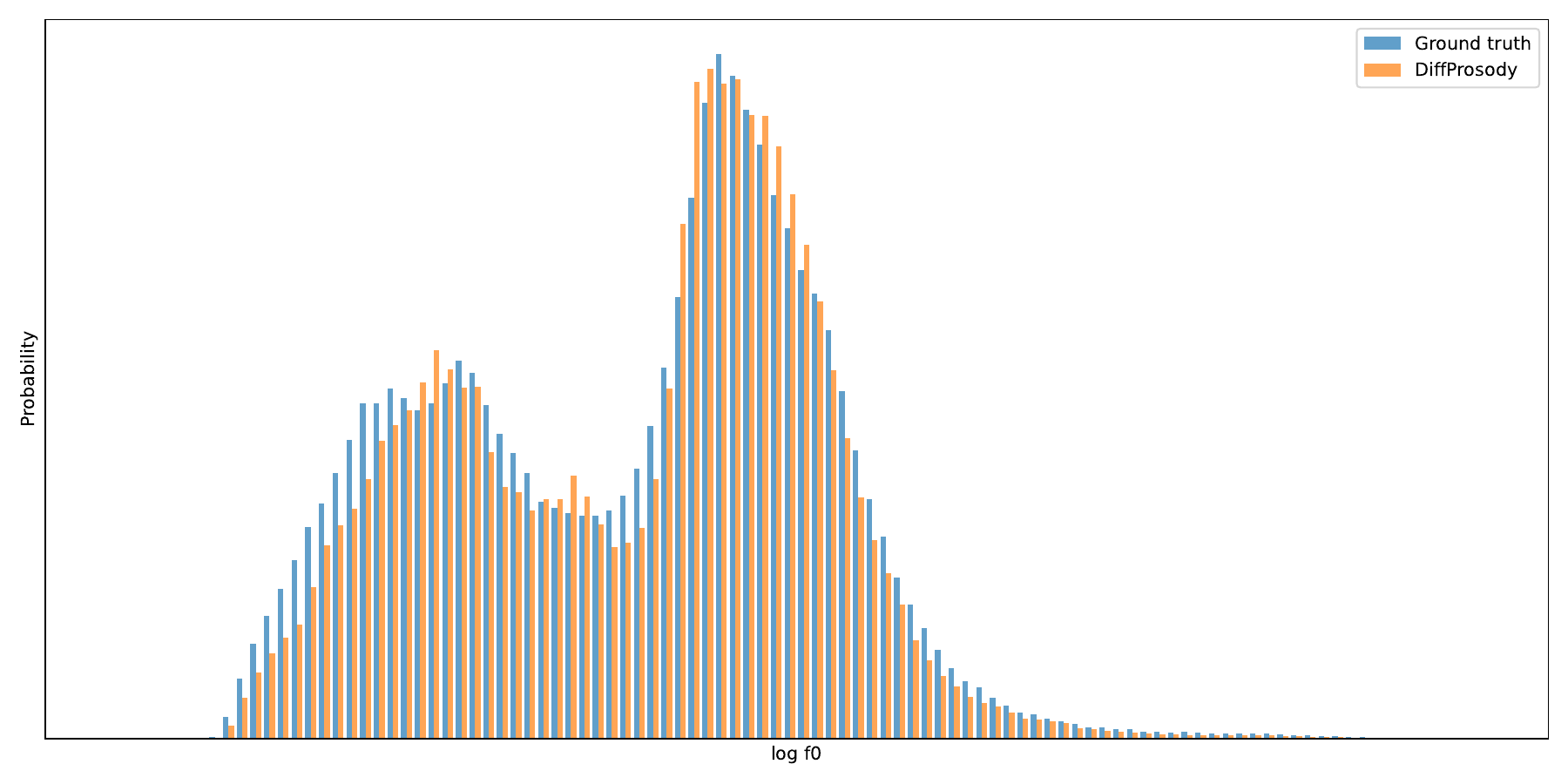}%
\label{fig_log_f0_pds}}

\caption{Histogram visualization of log f0, where the blue bars represent the GT distribution and orange bars represent the generated distribution. The distribution of the proposed model overlaps to a greater extent with the GT distribution than the other comparison models.}
\label{fig_f0_sim}
\centering

\captionsetup[subfloat]{labelfont=scriptsize,textfont=scriptsize}  % subfloat caption setting
\subfloat[FastSpeech2]{\includegraphics[width=0.3\columnwidth]{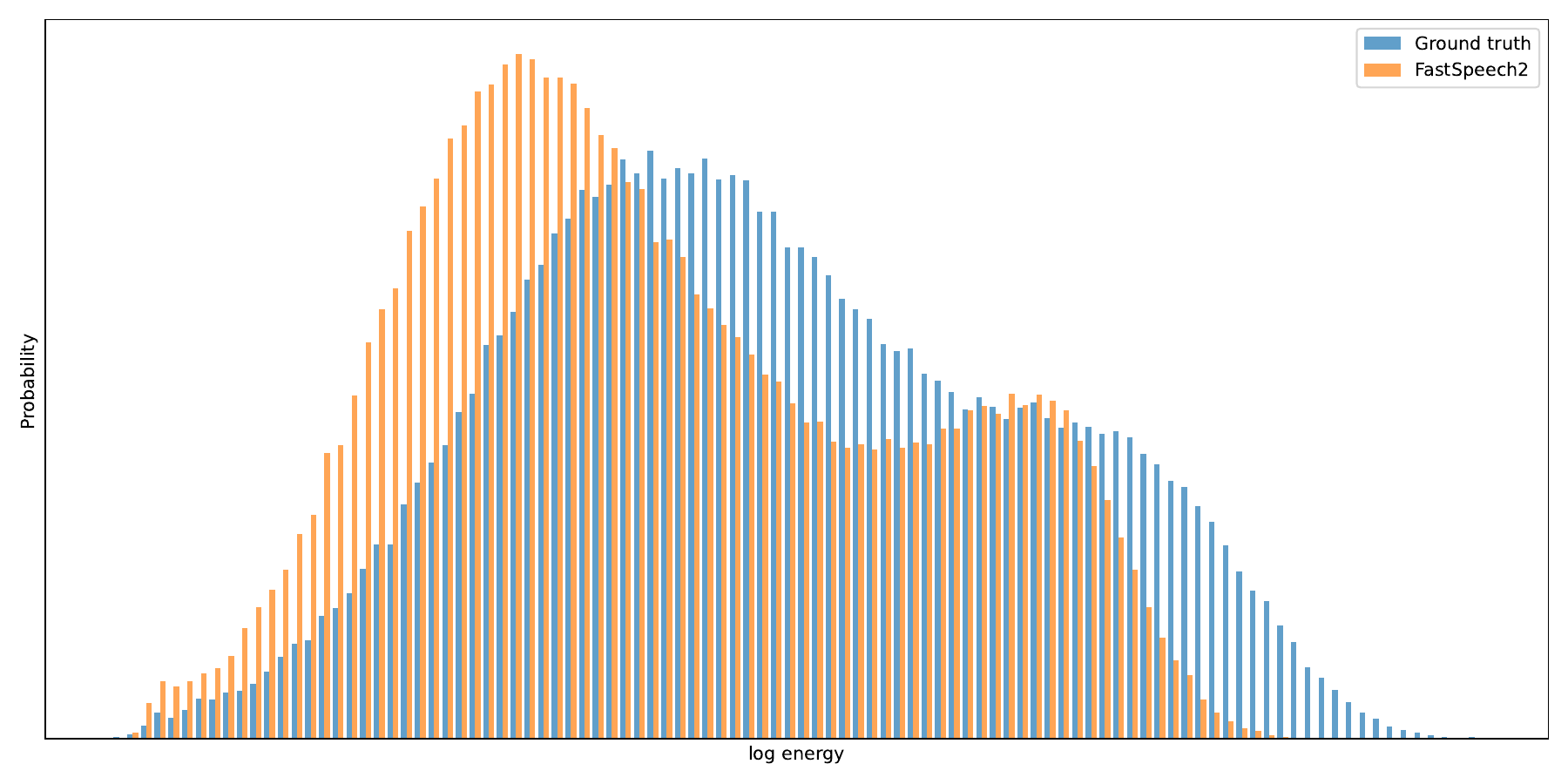}%
\label{fig_log_energy_fs2}}
\hspace{0.1cm}
\subfloat[ProsoSpeech]{\includegraphics[width=0.3\columnwidth]{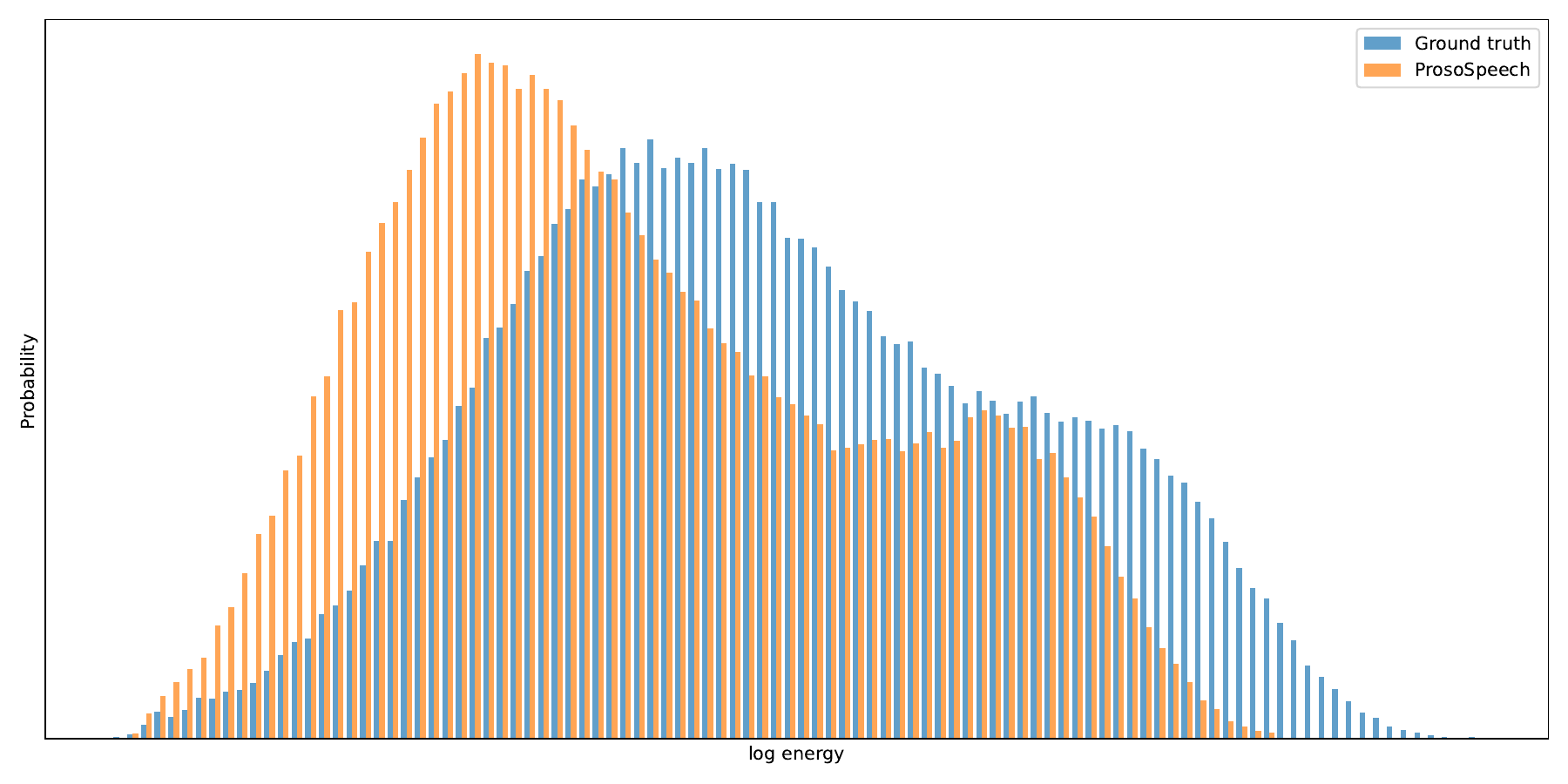}%
\label{fig_log_energy_prs}}
\hspace{0.1cm}
\subfloat[DiffProsody]{\includegraphics[width=0.3\columnwidth]{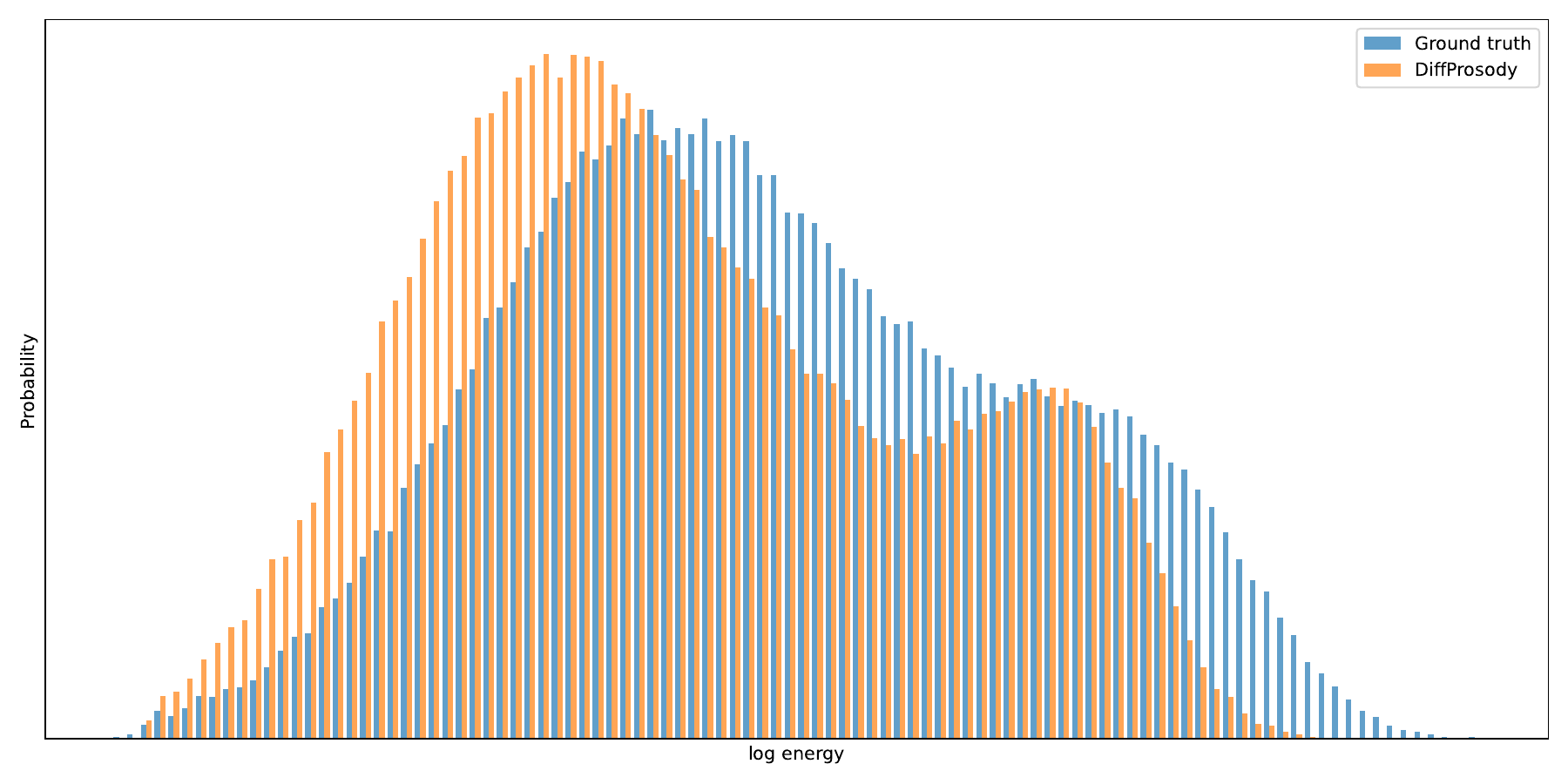}%
\label{fig_log_energy_pds}}
\caption{Histogram visualization of log energy, where the blue bars represent the GT distribution and orange bars represent the generated distribution. The proposed model’s distribution has a greater overlap with the GT distribution compared to the other models being compared.}
\label{fig_energy_sim}
\end{figure}

Figure \ref{fig_f0_sim} presents the histogram distribution for log f0 (pitch), and Figure \ref{fig_energy_sim} presents the histogram distribution for log energy. 
In both the figures, the blue bars represent the distribution of the GT features, and the orange bars represent the distribution of the generated features. The results indicate that the proposed model aligns more closely with the GT distribution than the other models. 
Table \ref{table:kld} lists the KL divergence values for comparison. ProsoSpeech exhibited a better performance in f0 than FastSpeech 2, but not in the case of energy. 
However, DiffProsody outperformed the comparison model in terms of f0 and energy.

\subsection{Latent prosody generation method}
In Table \ref{table:ablation}, DiffProsody (AR) is a model comprising the use of an AR prosody predictor. DiffProsody (DDPM) is a model comprising the use of DDPM (100 timesteps). DiffProsody is a model comprising the use of DDGAN (four timesteps). 
The AR predictor has the same structure as the prosody predictor in ProsoSpeech, but it does not use context encoders, and the denoising network in the DDPM has the same structure as generator $g_\theta$ in the DDGAN.
We also conducted a 7-point comparative mean opinion score (CMOS) evaluation to compare the latent prosody generation methods and measured the RTF to compare the generation speeds. 
DDGAN achieved $-$0.172 and $+$0.015 CMOS compared with AR and DDPM, respectively. 
The objective metrics showed that the DDPM and DDGAN outperformed the AR.
The DDGAN achieved results nearly identical to those of the DDPM for all the metrics. 
The experimental results showed that the diffusion models performed better than the AR models, as reported in\cite{ramesh2022hierarchical}.
Furthermore, DDGAN can generate high-quality prosody vectors such as DDPM in only four timesteps.
According to the RTF results, the DDGAN produces a 2.7$\times$ and 16$\times$ faster prosody than the AR and DDPM, respectively.

\begin{table}[!t]
  \caption{KL Divergence of log f0 and energy} % \vspace{-0.3cm}
  \centering
  \resizebox{0.9\columnwidth}{!}{
  \begin{tabular}{lccc}
    \toprule
    \textbf{Number of Mel-bins}     & \textbf{KL divergence $\log{f0}$} & \textbf{KL divergence $\log{energy}$} \\ \midrule
      FastSpeech 2          & 0.00574 & 0.02375     \\
      ProsoSpeech           & 0.00542 & 0.02657     \\
      DiffProsody           & \textbf{0.00343} & \textbf{0.01547}     \\ 
    \bottomrule
  \end{tabular}}  
  \label{table:kld} % \vspace{-0.6cm}
\end{table}

\begin{figure}[!t]
\centering
\captionsetup[subfloat]{labelfont=scriptsize,textfont=scriptsize}  % subfloat caption setting
\subfloat[WER]{\includegraphics[width=0.33\columnwidth]{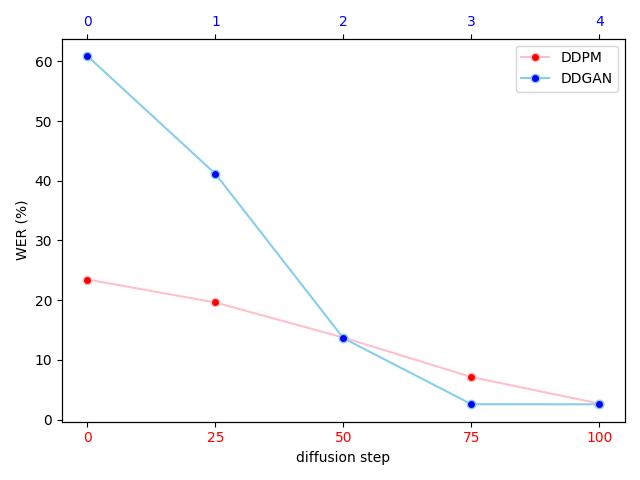}%
\label{fig_trace_w_wer}}
\subfloat[CER]{\includegraphics[width=0.33\columnwidth]{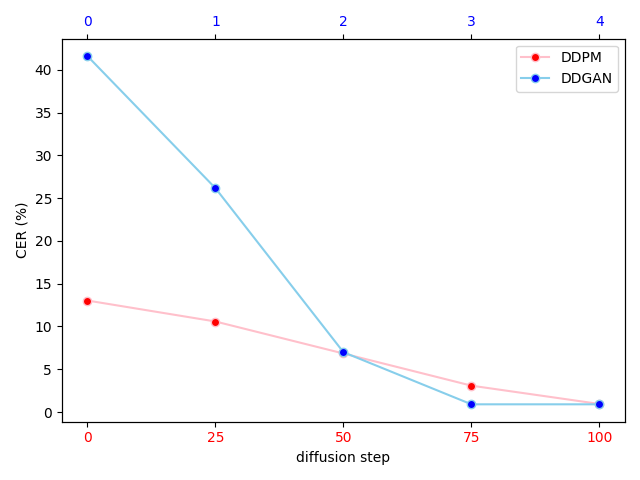}%
\label{fig_trace_w_cer}}
\subfloat[EER]{\includegraphics[width=0.33\columnwidth]{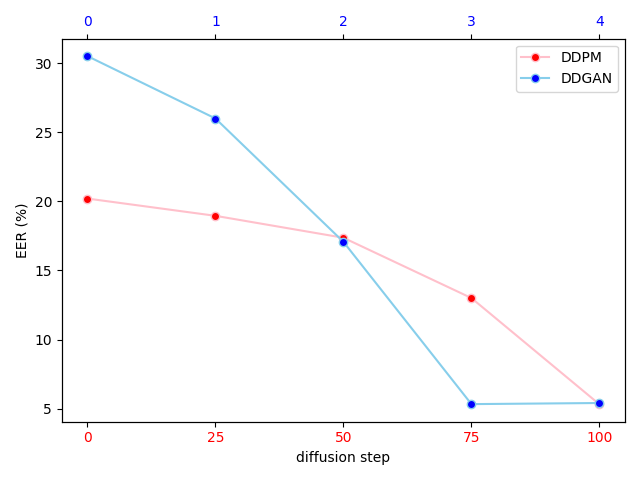}%
\label{fig_trace_w_eer}}
\caption{Comparison of objective evaluation results based on diffusion timesteps when using the DDPM and DDGAN framework in DLPG. The blue line is the result for the DDGAN and the red line is the result for the DDPM.}
\label{fig_trace}
\end{figure}

Figure \ref{fig_trace} presents the traces of the WER, CER, and EER results for the speech generated by the DLPG with the prosody vector generated by each denoising iteration. The red and blue lines represent the results obtained using the DDPM and DDGAN, respectively. We observed that the DDGAN has a larger error rate than the DDPM in the early stages and that the midpoint and final stages of each model have almost the same value. This implies that the DDGAN compresses the timesteps of the DDPM by modeling the denoising process as a multimodal non-Gaussian distribution. 

\begin{table*}[!ht]
  \caption{Objective and subjective evaluations of ablation study models} % \vspace{-0.3cm}
  \centering
  \resizebox{1.0\textwidth}{!}{
  \begin{tabular}{l|c|ccc|cccc|c|c}
    \toprule
    \textbf{Method}           & \textbf{CMOS} & \textbf{CER (\%)} ($\downarrow$) & \textbf{WER (\%)} ($\downarrow$) & \textbf{DDUR} ($\downarrow$) & \textbf{EER (\%)} ($\downarrow$) & \textbf{RMSE$_{period}$} ($\downarrow$) & \textbf{RMSE$_{period}$} ($\downarrow$) & \textbf{F1$_{v/uv}$} ($\uparrow$) & \textbf{RTF} ($\downarrow$) & \textbf{Params} \\ \midrule
      DiffProsody           & 0.0       & 0.90 & 2.55 & 0.295 & 5.404 & 54.60 & 0.475 & 0.6952 & 0.054 & 53M\\ 
      \midrule
      DiffProsody (AR)        & $-$0.172      & 1.41 & 3.73 & 0.308 & 6.947 & 65.63 & 0.486 & 0.6797 & 0.134 & 59M \\ 
      DiffProsody (DDPM)       & $+$0.015       & 0.96 & 2.76 & 0.282 & 5.263 & 54.04 & 0.477 & 0.6933 & 0.871 & 53M \\
      \midrule
      DiffProsody (w/o PCD)        & -      & 1.19 & 3.39 & 0.306 & 5.762 & 56.09 & 0.477 & 0.6921 & 0.054 & 53M \\ 
      DiffProsody (w/o VQ)        & -      & 1.41 & 3.73 & 0.308 & 7.306 & 62.03 & 0.482 & 0.6783 & 0.054 & 53M \\       
    \bottomrule
  \end{tabular}
   }
  \label{table:ablation} % \vspace{-0.6cm}
\end{table*}

\begin{table}[!t]
  \caption{CMOS results for prosody conditional discriminator} % \vspace{-0.3cm}
  \centering
  \resizebox{0.8\columnwidth}{!}{
  \begin{tabular}{lccc}
    \toprule
    \textbf{Method}           & \textbf{reference} & \textbf{CMOS} \\ \midrule
      DiffProsody           & \checkmark & 0.0 & \\
      DiffProsody (w/o PCD)      &  & $-$0.171 \\ 
      \midrule
      DiffProsody           & \checkmark & 0.0 & \\ 
      ProsoSpeech           &  & $-$0.183 \\
      \midrule
      DiffProsody (w/o PCD)      & \checkmark & 0.0 & \\
      ProsoSpeech           &  & $-$0.065 \\
    \bottomrule
  \end{tabular}}  
  \label{table:pcd} % \vspace{-0.6cm}
\end{table}

\subsection{Prosody conditional adversarial training}
We conducted CMOS to assess the effectiveness of PCD. The CMOS values were used to measure the degree of PCD preference in comparison with the reference model. Our evaluation involved the analysis of three different models: DiffProsody using PCD and a diffusion-based model, DiffProsody without PCD (referred to as DiffProsody (w/o PCD)) using a diffusion-based model, and ProsoSpeech using an AR prosody predictor without PCD. The results presented in Table \ref{table:pcd} clearly indicate that models comprising the use of the PCD are preferred over those trained without the PCD.
Furthermore, by comparing the CMOS results of DiffProsody (w/o PCD) and ProsoSpeech, we can infer that the diffusion-based method is preferable to the method employed by ProsoSpeech. To provide a more comprehensive analysis, we also incorporated the objective metrics for DiffProsody (w/o PCD) in Table \ref{table:ablation}. These objective metric results demonstrate that DiffProsody outperformed DiffProsody (w/o PCD) in all aspects. However, DiffProsody (w/o PCD) still received a better objective evaluation than ProsoSpeech.
In addition, when comparing DiffProsody (AR) with ProsoSpeech, DiffProsody (AR) consistently achieved higher scores on the majority of the objective evaluations. These findings validate that both PCD and DLPG significantly improve the model performance.

\subsection{Prosody encoder evaluation}
In this section, the effectiveness of the prosody encoder is evaluated. We focus on two main aspects: the impact of the number of Mel-bins used in the reference Mel-spectrogram and the role of the vector quantization layer. Detailed results of these evaluations are presented in the following subsections.

\begin{figure*}[!t] % \vspace{-0.8cm}
  \centering
  \includegraphics[width=0.95\textwidth]{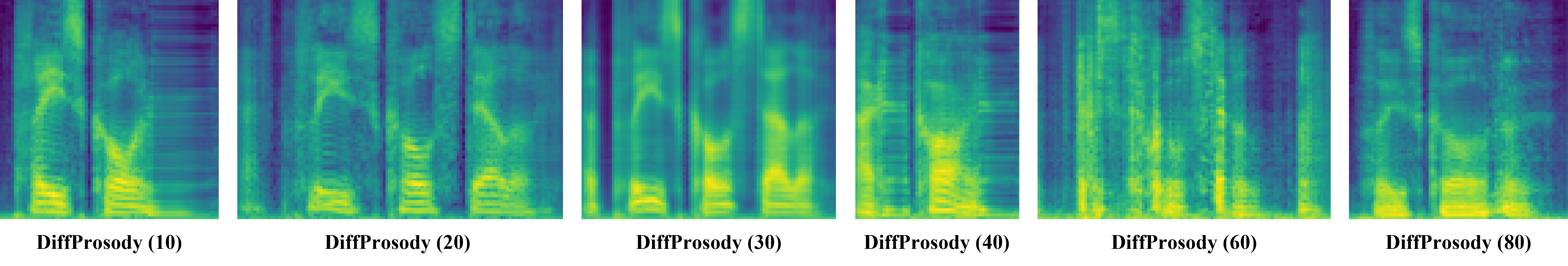}
  \caption{Comparison of Mel-spectrograms at iteration 0 (noise) in diffusion with DiffProsody trained on the number of bins $N$ in the Mel spectrogram. The Mel-spectrogram exhibits a gradual collapse as $N$ increases. This occurs when disentangling has failed, and the prosody vector is responsible for the linguistic information.} % \vspace{-0.65cm}
\label{fig_bins}
\end{figure*}

\subsubsection{Impact of the number of Mel-bins}
We conducted an experiment to compare the performance of DiffProsody when trained using various numbers of Mel-bins. The results, including the EER, CER, and WER scores, are presented in Table \ref{table:mel_bins}. We examined the performance of 10, 20, 30, 40, 60, and 80 (full-band) Mel-bins. The findings indicated that as the number of bins exceeded 20 (baseline), the EER tended to increase, while no significant difference was observed in terms of the CER and WER. It should be noted that the model with 10 bins outperformed the baseline (20 bins) in terms of the WER but yielded higher EER results.
Figure \ref{fig_bins} illustrates the Mel-spectrogram at the initial iteration (noise) during the diffusion timestep of the DLPG, which was trained using various numbers of Mel-bins. 
As the number of Mel-bins used for the training was increased, the results of the Mel-spectrogram progressively smoothened and eventually collapsed. 
This phenomenon occurs because the large amount of information in the reference forces the model to reconstruct the Mel-spectrogram by leveraging the prosody vectors. 
Through this experiment, we found that, as $N$ increases, linguistic information becomes increasingly entangled. Consequently, it is reasonable to employ 20 Mel-bins as the input to the prosody encoder for realizing effective prosody modeling.

\begin{figure}[!t]
\centering
\captionsetup[subfloat]{labelfont=scriptsize,textfont=scriptsize}  % subfloat caption setting
\subfloat[DiffProsody (w/o VQ)]{\includegraphics[width=1.0\columnwidth]{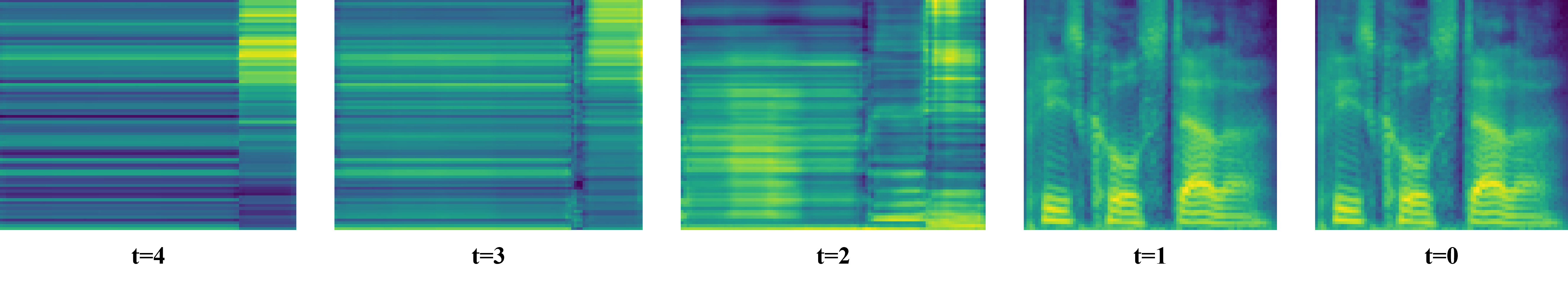}%
\label{fig_no_vq}}

\subfloat[DiffProsody]{\includegraphics[width=1.0\columnwidth]{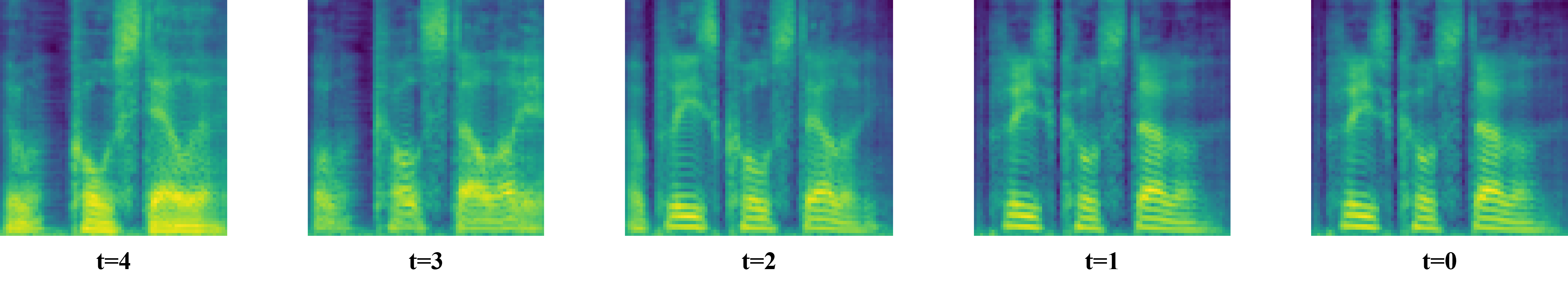}%
\label{fig_vq}}
\caption{Visualization of a Mel-spectrogram trace synthesized with latent generation for each diffusion timestep. (a) DiffProsody (w/o VQ) trained without vector quantization in prosody encoder; (b) DiffProsody. (a) shows that,  in contrast (b), the early stages of diffusion produce a completely collapsed Mel-spectrogram. This is caused by the inclusion of linguistic information in the prosody vector.}
\label{trace_mel}
\end{figure}

\subsubsection{Vector quantization layer analysis}
Figure \ref{trace_mel} presents a Mel-spectrogram trace synthesized using the prosody vector generated for each diffusion step of the DLPG. We compared two versions of DiffProsody: one trained without the vector quantization layer (Figure \ref{trace_mel}a; defined as DiffProsody (w/o VQ)) and the other trained with the vector quantization layer (Figure \ref{trace_mel}b; defined as DiffProsody).
In the case of DiffProsody (w/o VQ), the early steps exhibited a completely distorted Mel-spectrogram, but there was a significant recovery in the middle steps. Conversely, the initial step of DiffProsody exhibits a smoothed but slightly distorted Mel-spectrogram that gradually returns to its original state over the subsequent steps. 
This phenomenon is also related to prosody disentangling, and we confirmed that the prosody disentangling failed in DiffProsody (w/o VQ).
This experiment demonstrated that vector quantization plays a crucial role in effective prosody disentangling.
The last row of Table \ref{table:ablation} presents the objective evaluation results for DiffProsody (w/o VQ). DiffProsody (w/o VQ) performed worse for all the objective measurements. These results provide an objective assessment of how the failure to properly disentangle prosody affects the overall performance of the system.

\begin{table}[!t]
  \caption{Objective evaluation based on the number of Mel-bins} % \vspace{-0.3cm}
  \centering
  \resizebox{0.75\columnwidth}{!}{
  \begin{tabular}{lccc}
    \toprule
    \textbf{Number of Mel-bins}     & \textbf{EER} & \textbf{CER} & \textbf{WER} \\ \midrule
      DiffProsody (10)               & 5.71 & 0.95 & \textbf{2.49}    \\
      DiffProsody (20)               & \textbf{5.40} & \textbf{0.90} & 2.55    \\
      DiffProsody (30)               & 6.26 & 1.0 & 2.58    \\ 
      DiffProsody (40)               & 6.56 & 0.98 & 2.59    \\ 
      DiffProsody (60)               & 6.20 & 0.97 & 2.52    \\ 
      DiffProsody (80)               & 7.11 & 0.98 & 2.67    \\
    \bottomrule
  \end{tabular}}  
  \label{table:mel_bins} % \vspace{-0.6cm}
\end{table}

\section{Conclusion}
In this study, a novel technique called DiffProsody is proposed, the aims of which is to synthesize high-quality expressive speech. Through prosody conditional adversarial training, we observed significant improvements in speech quality with a more pronounced display of expressive prosody. In addition, our DLPG successfully generated expressive prosody. Our proposed method outperformed comparative models in terms of producing accurate and expressive prosody, as evidenced by the prosody evaluation metrics. Moreover, our method demonstrated superior accuracy in pronunciation, as indicated by the CER and WER evaluations. The KL divergence and histogram analysis further support the claim that DiffProsody yields a more accurate prosody distribution than the other models. Furthermore, we successfully reduced the sampling speed while maintaining the expected performance by introducing DDGAN. 

\section{Future works}
Despite the importance of vector quantization for disentangling, it has been observed that this approach can negatively affect model performance. 
This problem is expected to be addressed with the introduction of methods such as residual vector quantizers\cite{9625818, 10158503, hwang2023hiddensinger}.
Moreover, we acknowledge the limitations of attempting to model prosody using the TTS dataset. 
It has been suggested that using a language model pre-trained on a large dataset, such as HuBERT\cite{9585401}, could result in significant improvements. 
In the future, we plan to extend the latent diffusion method to include controllable emotional prosody modeling\cite{9747098}.

\bibliographystyle{IEEEtran}
\bibliography{main}
\begin{IEEEbiography}[{\includegraphics[width=1in,height=1.25in,clip,keepaspectratio]{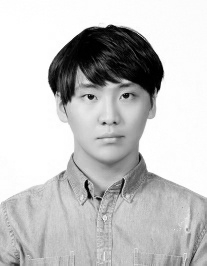}}]{Hyung-Seok~Oh} 
received the B.S. degree in Computer Science and Engineering from Konkuk University, Seoul, South Korea, in 2021. 
He is currently working toward the integrated master’s and Ph.D. degree with the Department of Artificial Intelligence, Korea University, Seoul, South Korea. 
His research interests include artificial intelligence and audio signal processing.
\end{IEEEbiography}
\begin{IEEEbiography}[{\includegraphics[width=1in,height=1.25in,clip,keepaspectratio]{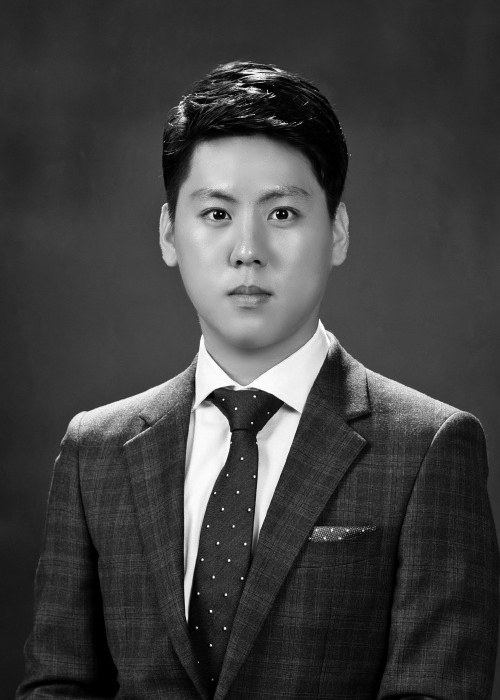}}]{Sang-Hoon~Lee}
received the B.S. degree in life science from Dongguk University, Seoul, South Korea, in 2016 and received the Ph.D. degree in Brain and Cognitive Engineering from Korea University, Seoul, South Korea, in 2023. He is currently a postdoctoral researcher in AI Research Center, Korea University, Seoul, South Korea. His current research interests include artificial intelligence and audio signal processing.
\end{IEEEbiography}
\begin{IEEEbiography}[{\includegraphics[width=1in,height=1.25in,clip,keepaspectratio]{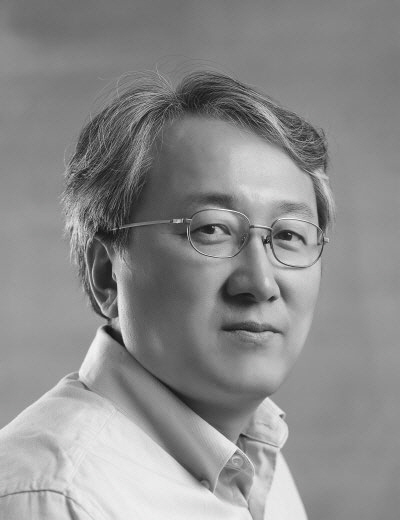}}]{Seong-Whan~Lee}
(Fellow, IEEE) received the B.S. degree in computer science and statistics from Seoul National University, South Korea, in 1984, and the M.S. and Ph.D. degrees in computer science from the Korea Advanced Institute of Science and Technology, South Korea, in 1986 and 1989, respectively. He is currently the Head of the Department of Artificial Intelligence, Korea University, Seoul. His current research interests include artificial intelligence, pattern recognition, and brain engineering. He is a Fellow of the International Association of Pattern Recognition (IAPR) and the Korea Academy of Science and Technology.
\end{IEEEbiography}
\end{document}